\documentclass[11pt,a4paper]{article}
\pdfoutput=2
\usepackage{jheppub} 
\usepackage{graphicx}
\usepackage{dcolumn}
\usepackage{bm}
\usepackage{adjustbox}
\usepackage{multirow}
\usepackage{slashed}
\usepackage{hhline}
\usepackage{mathrsfs}
\usepackage[utf8]{inputenc}
\usepackage{tabularx,booktabs,caption}
\usepackage[colorlinks=false,
urlcolor=cambridgeblue,
linkcolor=blue,
citecolor=cambridgeblue,
linktocpage=true,
pdfproducer=medialab,
pdfa=true,
anchorcolor=blue]{hyperref}%

\usepackage{cancel}
\usepackage{amssymb}
\usepackage{textcomp}
\usepackage{amsmath}
\usepackage{mathrsfs}
\usepackage{subfigure}

\usepackage{empheq}
\usepackage{bm}
\usepackage{framed}
\usepackage{pdfpages}
\usepackage{subfigure}

\usepackage{float}
\usepackage{xcolor}

\newcommand{\gsim}{\gtrsim}
\newcommand{\lsim}{\lesssim}

\newcommand{\lf}{\left(}
\newcommand{\ri}{\right)}

\newcommand{\nn}{\nonumber}

\newcommand{\rr}{{\gamma\gamma}}
\renewcommand{\lg}{\mathscr{L}}

\newcommand{\mcd}{\mathcal{D}}
\newcommand{\mcf}{\mathcal{F}}

\newcommand{\mcq}{\mathcal{Q}}

\newcommand{\mcu}{\mathcal{U}}

\newcommand{\bmcq}{\overline{\mathcal{Q}}}

\newcommand{\br}{\mathcal{B}}
\newcommand{\hc}{{\rm H.c.}}

\newcommand{\sm}{{\rm SM}}
\newcommand{\tot}{{\rm tot}}
\newcommand{\np}{{\rm NP}}

\newcommand{\fb}{{\;{\rm fb}}}
\newcommand{\ifb}{{\;{\rm fb}^{-1}}}

\newcommand{\gev}{{\;\,{\rm GeV}}}
\newcommand{\tev}{{\;\,{\rm TeV}}}

\newcommand{\beq}{\begin{equation}}
\newcommand{\eeq}{\end{equation}}
\newcommand{\bea}{\begin{eqnarray}}
\newcommand{\eea}{\end{eqnarray}}
\newcommand{\barr}{\begin{array}}
\newcommand{\earr}{\end{array}}
\newcommand{\bc}{\begin{center}}
\newcommand{\ec}{\end{center}}
\newcommand{\bit}{\begin{itemize}}
\newcommand{\eit}{\end{itemize}}
\newcommand{\ben}{\begin{enumerate}}
\newcommand{\een}{\end{enumerate}}

\newcommand{\al}{\alpha}
\newcommand{\bt}{\beta}

\newcommand{\dt}{\delta}
\newcommand{\Dt}{\Delta}

\newcommand{\sg}{\sigma}

\newcommand{\kp}{\kappa}

\newcommand{\gm}{\gamma}
\newcommand{\Gm}{\Gamma}
\newcommand{\lm}{\lambda}

\newcommand{\mhh}{M_{H}}

\newcommand{\ca}{c_\alpha}
\newcommand{\sa}{s_\alpha}

\newcommand{\tb}{t_\beta}

\newcommand{\cb}{c_\beta}
\renewcommand{\sb}{s_\beta}

\newcommand{\cba}{c_{\beta-\alpha}}
\newcommand{\sba}{s_{\beta-\alpha}}
\newcommand{\sth}      {s_\theta}
\newcommand{\cth}      {c_\theta}

\newcommand{\mh}{m_{h}}

\newcommand{\ttau}     {{\tau\tau}} 

\newcommand{\ttop}      {{t\bar{t}}}

\newcommand{\bb}      {{b \bar{b}}}

\vspace*{18pt}

\title{Disentangling new physics effects 
on non-resonant Higgs boson pair production from gluon fusion}

\author[a,b,c]{Kingman Cheung,}
\emailAdd{cheung@phys.nthu.edu.tw}

\author[a]{Adil Jueid,}
\emailAdd{adil.hep@gmail.com}

\author[d]{Chih-Ting Lu,}
\emailAdd{timluyu@gmail.com}

\author[a]{Jeonghyeon Song,}
\emailAdd{jhsong@konkuk.ac.kr}

\author[e]{and Yeo Woong Yoon}
\emailAdd{ywyoon@kias.re.kr}

\affiliation[a]{Department of Physics, Konkuk University, Seoul 05029, Republic of Korea}
\affiliation[b]{Physics Division, National Center for Theoretical Sciences, Hsinchu, Taiwan}
\affiliation[c]{Department of Physics, National Tsing Hwa University, Hsinchu 300, Taiwan}
\affiliation[d]{School of Physics, KIAS, Seoul 130-722, Republic of Korea}
\affiliation[e]{School of Liberal Arts, Seoul-Tech, Seoul 139-743, Republic of Korea}

\date{\today}

\abstract{
There are two kinds of new physics effects
on non-resonant di-Higgs process from gluon fusion,
non-SM Higgs trilinear self-coupling $\lambda_{hhh}$ or
new colored particles running in the loop.
With the aim of disentangling different new physics contributions,
we study their characteristics in the kinematic distributions.
Assuming that the total cross section 
is observed to be about three times as large as the SM expectation,
we consider the cases of
$\lambda_{hhh}/\lambda_{hhh}^{\rm SM} =-0.5, 5.5$
as well as a new physics model with heavy vectorlike quarks 
in a type-II two Higgs double model, called the VLQ-2HDM.
A reasonable benchmark point is suggested in the exact wrong-sign limit,
where the opposite sign between 
the up-type VLQ and down-type VLQ couplings to the Higgs boson
causes the cancellation of their contributions to the single-Higgs production from gluon fusion.
Because of the threshold effects from the heavy VLQs in the loop,
the VLQ-2HDM accommodates the bumps in the distributions of the invariant mass
of the Higgs boson pair ($M_{hh}$) and the transverse momentum of a Higgs boson ($p_T^h$).
The positions of two bumps are closely related as $ M_{hh} \simeq 2 M_{\rm VLQ}$
and $p_T^h \simeq M_{\rm VLQ}$.
In addition, the bumps located at the heavy VLQ mass
naturally lift up the $M_{hh}$ and $p_T^h$ distributions into high-mass and
high-$p_T^h$ regions.
On the other hand,
the non-SM Higgs trilinear coupling cases 
have the distributions shift into low $ M_{hh}$ and $p_T^h$ regions.
Therefore, the kinematic region with high $M_{hh}$ and high $p_T^h$
will be a smoking-gun signal for the VLQ-2HDM.
Full HL-LHC simulations for the di-Higgs signals are also performed,
confirming that the $b \bar{b} b \bar{b}$ final state can distinguish
the VLQ-2HDM.
}

\begin{document}
\maketitle

\section{Introduction}
\label{sec:intro}

In particle physics,
a great step forward 
in knowledge or model building has always been realized by the observation of a new interaction vertex.
The discovery of the Higgs boson at the LHC by the ATLAS and CMS collaborations~\cite{Aad:2012tfa, Chatrchyan:2012xdj}
was also based on the measurement of the couplings of a new scalar boson to vector bosons 
and the third generation fermions. 
Even though all of the experimental results conform to the phenomenology of the standard model (SM) Higgs boson~\cite{Aad:2019mbh},
proving the converse,
the discovery of \textit{the SM Higgs boson},
requires additional and unprecedented steps, 
measuring the Higgs trilinear and quartic self-couplings as well as the couplings 
to the first and second generation fermions.
At the high-luminosity LHC (HL-LHC),
two couplings among them are expected to be observed, 
the Higgs coupling to a muon pair
and the Higgs trilinear self-coupling $\lm_{hhh}$~\cite{CMS:2018qgz,deBlas:2019rxi}.
As the Higgs self-interaction is the key 
to understand electroweak symmetry breaking, vacuum stability, 
and electroweak phase transition,
the new physics (NP) hunters
rely more on $\lm_{hhh}$, 
which is to be probed via Higgs boson pair production  
at the LHC,
simply called the di-Higgs process~\cite{Plehn:2005nk,deBlas:2019rxi,Li:2019uyy,Cao:2016zob,Cao:2015oaa}.

The major production channel for the di-Higgs process
is the gluon-gluon fusion,
which receives the contributions from the triangle and box diagrams through the top and bottom quarks in the SM~\cite{Glover:1987nx,Plehn:1996wb}. 
The triangle diagram is mediated by the Higgs boson in $s$-channel,
providing the connection to $\lm_{hhh}$.
There are three main ways to accommodate NP effects 
on $gg \to hh$.
The first is 
the resonant production of the Higgs 
boson pair through a new scalar boson or the spin-2 Kaluza-Klein graviton in the Randall-Sundrum model~\cite{Djouadi:1999rca,Dolan:2012ac,Chen:2014ask,No:2013wsa,Alves:2018oct,Adhikary:2017jtu}.
The second is non-SM Higgs trilinear self-coupling~\cite{Asakawa:2010xj,Baglio:2012np,Barger:2013jfa,Slawinska:2014vpa}, 
parameterized by the Higgs coupling modifier $\kp_\lm \equiv \lm_{hhh}/\lm_{hhh}^\sm$.
The third is to introduce new colored particles in the triangle and box diagrams~\cite{Batell:2015koa,Huang:2017nnw,Kribs:2012kz,Lee:2004me,Dawson:2015oha,Cao:2013si,Han:2013sga,Raju:2020hpe,Alves:2019igs}.\footnote{There is 
another interesting way 
through model-independent dimension-six effective
operators,
which yields the derivative cubic Higgs coupling~\cite{He:2015spf,Alloul:2013naa,Pierce:2006dh,Buchalla:2018yce}.}

Since resonant Higgs boson pair production can be identified 
through a peak in the distribution of the invariant mass of the Higgs boson pair,
experimentalists usually present the di-Higgs results
in two modes, non-resonant and resonant ones~\cite{Aad:2019uzh,Sirunyan:2018two}.
And the result of non-resonant mode 
is usually translated into the limit on $\kp_\lm$ such that
the latest one is $-5.0 < \kp_\lm < 12.0$ at 95\% confidence level~\cite{Aad:2019uzh}.
Even though this is a reasonable choice at the moment
with the very small number of signal events, 
we point out that 
non-resonant NP effects have another source of
new colored particles.
Expecting higher discovery potential of the di-Higgs process in the future,
the key question is how to distinguish different non-resonant NP effects
if we observe considerably a large di-Higgs production cross section.

A unique feature of heavy particles running in the loop
is their threshold effect.
One good example is the top quark threshold contributions
to the gluon fusion production of a photon pair at the LHC~\cite{Jain:2016kai},
which appears as a bump 
around $M_\rr \simeq 2 m_t$. 
Any new heavy particle $\mcf$ in the loop,
if enhancing the di-Higgs process,
would yield a similar bump structure
in the $M_{hh}$ distribution~\cite{Dawson:2015oha}.
Simply with non-SM Higgs trilinear self-coupling,
we cannot accommodate this irregular structure of the threshold origin.
In addition, 
a naive parton level kinematics
in the limit of $M_\mcf \gg m_h$
predicts that the events corresponding to $M_{hh} \simeq 2 M_\mcf$
prefer
$p_T^h \simeq M_\mcf$
when the longitudinal motion is soft.
A correlation in the $M_{hh}$ and $p_T^h $ distributions
can be a smoking-gun signal for the new colored particles 
in the loop of the di-Higgs process.
This is our driving motivation.

We shall begin with the assumption 
that the total production cross section of the di-Higgs process
would be about three times as large as the SM expectation,
i.e., $\sg_\np/\sg_\sm (gg\to hh) \simeq 3$.
For the new colored particles, 
we consider the vectorlike quarks (VLQs) with a mass around the electroweak scale,
which not only appear in many new physics models~\cite{Cacciapaglia:2011fx, Aguilar-Saavedra:2013qpa, Ellis:2014dza, Angelescu:2015uiz, Arhrib:2016rlj, Barducci:2017xtw, Cacciapaglia:2017gzh, Arhrib:2018pdi, Cacciapaglia:2018qep, Song:2019aav,Randall:1999ee,Cheng:1999bg,ArkaniHamed:2002qy,Han:2003gf,Cheng:2005as,Kang:2007ib}
but also fit well with the Higgs precision data~\cite{Anastasiou:2011qw,Anastasiou:2016cez}.
A crucial factor here
is the correlation between the single-Higgs and di-Higgs production rates,
because the same triangle diagrams 
from the VLQs in the di-Higgs process occur in the single-Higgs process.
The constraint from the observed single-Higgs production
is too strong to allow $\sg_\np/\sg_\sm (gg\to hh) \simeq 3$
if the Higgs boson couplings to the VLQs
are the SM-like.
We shall show that 
this correlation can be broken by extending the Higgs sector 
into the type-II two Higgs doublet model (2HDM)~\cite{Branco:2011iw}
in the exact wrong-sign limit~\cite{Ferreira:2014naa,Ferreira:2014dya,Biswas:2015zgk,Kang:2018jem,Das:2017mnu}:
the model is to be called the VLQ-2HDM.
A full analytic calculation of the form factors in the VLQ-2HDM
is also required in order to properly accommodate the bump structures
in the  $M_{hh}$
and $p_T^h $ distributions, which we will show in subsequent sections.

As one of the most challenging and significant processes to observe at the LHC,
the di-Higgs process has been intensively studied at the state-of-art level.
Theoretically the total production cross section was calculated 
at next-to-next-to-next-to-leading order (N$^3$LO) in the infinite top-quark mass limit 
and the next-to-leading order (NLO) with full top-quark mass dependence~\cite{Chen:2019fhs,Borowka:2016ehy,Borowka:2016ypz,Borowka:2017idc,Heinrich:2019bkc}.
The search strategies to maximize the discovery sensitivity
have been suggested for different decay channels 
such as $b\bar{b}b\bar{b}$~\cite{Wardrope:2014kya}, 
$\bb\rr$~\cite{Chang:2018uwu,Yao:2013ika},
and
$b\bar{b}WW^{(*)}$~\cite{Papaefstathiou:2012qe,Kim:2018cxf}.
On the experimental side,
the ATLAS~\cite{Aad:2015xja,Aad:2019uzh} and CMS collaborations~\cite{Sirunyan:2018two}
have performed the search,
in different final states such as $b\bar{b}b\bar{b}$~\cite{Aad:2015uka,Aaboud:2016xco,Aaboud:2018knk,Sirunyan:2017isc,Sirunyan:2018tki}, 
$b\bar{b}WW^{(*)}$~\cite{Aad:2019yxi,Sirunyan:2019quj}, $b\bar{b}\tau^+ \tau^-$~\cite{Aaboud:2018sfw}, 
$b\bar{b}\gamma\gamma$~\cite{Aad:2014yja,Aaboud:2016xco,Aaboud:2018ftw,Sirunyan:2018iwt}, 
$\gamma\gamma WW^{(*)}$~\cite{Aaboud:2018ewm,Aaboud:2018zhh}, and 
$WW^{(*)}WW^{(*)}$~\cite{Aaboud:2018ksn}.
In view of these circumstances,
a full collider simulation of the signal
is inevitable to claim a new method for disentangling non-resonant NP effects on the di-Higgs process.
We shall perform the HL-LHC simulation 
in the $\bb\bb$ and $\bb\rr$ final states
for the NP signals of
the VLQ-2HDM, $\kp_\lm=-0.5$, and $\kp_\lm=5.5$,
all of which yield $\sg_\np/\sg_\sm (gg\to hh)\simeq 3$.
It will be shown that
the correlated pattern in the high $p_T^h$ and high $M_{hh}$ regions
for the $\bb\bb$ final state
is of great use to distinguish the VLQ-2HDM from the $\kp_\lm \neq 1$ models.

The paper is organized in the following way.
In Sec.~\ref{sec:hh},
we begin with summarizing the characteristics of Higgs boson pair production
from gluon fusion.
Focusing on the non-resonant case,
we parameterize the NP effects and motivate our model,
the VLQ-2HDM.
In Sec.~\ref{sec:model},
we briefly review the VLQ-2HDM
and suggest an ansatz 
for vanishing Peskin-Takeuchi parameter $\hat{T}$~\cite{Peskin:1991sw}.
In Sec.~\ref{sec:characteristics},
we present the parton-level study of the VLQ-2HDM effects on the di-Higgs process,
including the full analytic calculation of the form factors from new VLQs.
For a benchmark point in the exact wrong-sign limit,
we show the differences among different NP models
in the kinematic distributions of $p_T^h$ and $M_{hh}$.
Section \ref{sec:simulation} deals with the full HL-LHC simulations
of three NP models and the SM in the $\bb\bb$ and $\bb\rr$ final states, 
focusing on the double differential cross sections.
Section \ref{sec:conclusion} contains our conclusions.

\section{Non-resonant di-Higgs production from gluon fusion}
\label{sec:hh}

\begin{figure}[!h] \centering
\includegraphics[width=\textwidth]{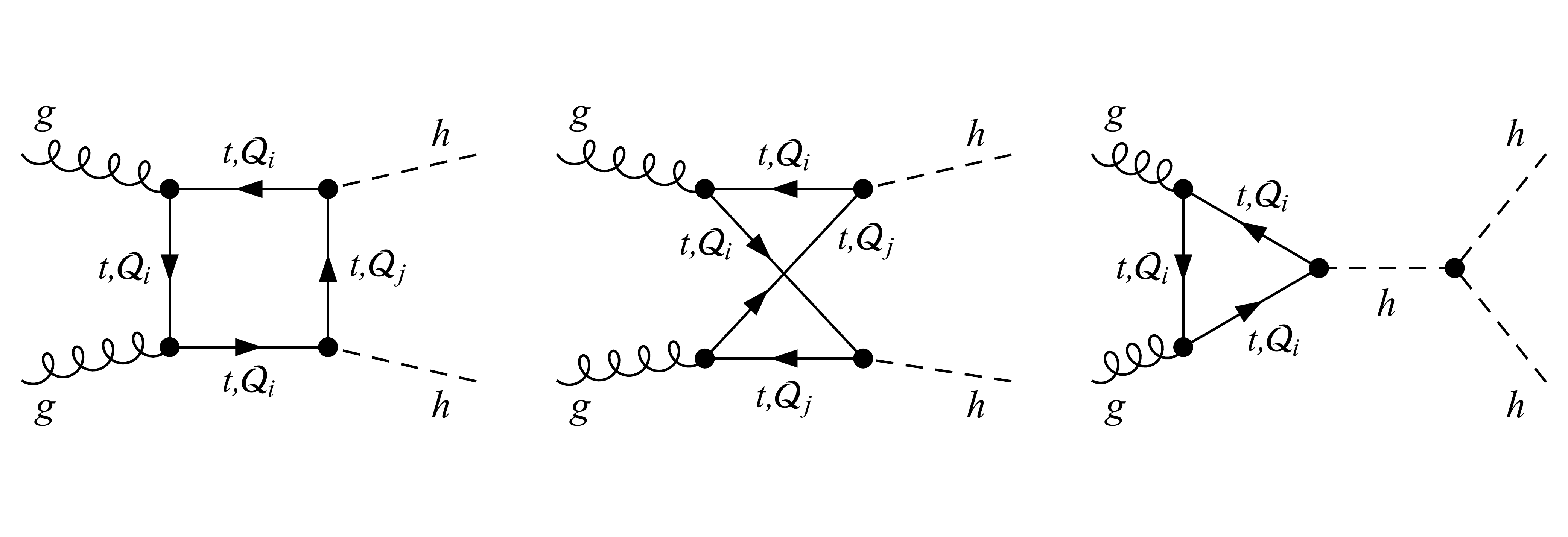}
\caption{\baselineskip 3.5ex
\label{fig:Feynman}
Representative Feynman diagrams for the di-Higgs process
via gluon-gluon fusion at the LHC. 
In addition to the SM top and bottom quarks,
new VLQs $\lf \mcq_i=\mcu_{1,2},\mcd_{1,2} \ri$
also contribute to the triangle and box diagrams.
}
\end{figure}

Gluon-gluon fusion production of a pair of Higgs bosons is
a loop-induced process
from two types of Feynman diagrams, 
triangle and box diagrams: see Fig.~\ref{fig:Feynman}.
In the SM, the top quark makes major contribution
to the process in both diagrams.
Since the triangle diagram is solely mediated by the Higgs boson
in $s$-channel, the Higgs trilinear coupling can be probed.
The partonic differential cross section to leading order
is~\cite{Plehn:1996wb}
\begin{eqnarray}
\label{eq:diffXShh:SM}
\frac{d\hat \sigma^\sm(gg\to hh) }{d \hat t} =
\frac{G_F^2 \alpha_s^2}{256 (2\pi)^3}
\left[ \left| \,
 \lambda_{hhh}^\sm \frac{ v} { \hat s - m_h^2  + i m_h \Gm_h} \, F_\triangle + F_\Box \right|^2 
+\left|  G_\Box \right|^2  \right]\,,
\end{eqnarray}
where $\lm_{hhh}^\sm\lf ={3m_h^2}/{v}\ri $ is the Higgs trilinear self-coupling,
and the expressions for $F_\triangle$, $F_\Box$, and $G_\Box$
 are referred to 
Ref.~\cite{Plehn:1996wb}.
In the low-energy theorem (LET) where $m_Q^2 \gg \hat{s}-m_h^2$, 
the form factors are simplified as
\bea
\label{eq:LET:SM}
F_\triangle^{\rm LET}  \simeq \frac{2}{3} ,\quad 
F_\Box^{\rm LET} \simeq -\frac{2}{3} ,\quad
G_\Box^{\rm LET} \simeq 0,
\eea
which clearly show
the destructive interference between the
triangle and box diagrams.
Special attention is required when using Eq.~(\ref{eq:LET:SM}).
Although they are useful in estimating the total production cross section,
the kinematic distributions based on the approximated form factors
are significantly different from the exact calculations,
especially in the high $p_T$ region~\cite{Dawson:2012mk}.

For illustrative purpose,
we assume that the di-Higgs process 
is observed at the HL-LHC
with the total cross section about three times
as large as the SM prediction:
\bea
\label{eq:3sigma}
\left. 
\frac{\sigma (gg\to hh)_{\rm NP}}{ \sigma (gg\to hh)_{\rm SM}} 
\right|_{14\,\rm TeV}  \simeq 3.
\eea
We further suppose that the possibility of resonant Higgs boson pair production
is ruled out from the study of the invariant-mass distribution
of two Higgs bosons (via e.g. $ h h \to b\bar b \gamma\gamma$
\cite{Chang:2018uwu}).
For non-resonant sources of Eq.~(\ref{eq:3sigma}), 
we consider the following two kinds of NP effects:
\begin{description}
\item[(i)] $\;\kp_\lm=-0.5,~ 5.5$;
\item[(ii)] new VLQs.
\end{description}
Two NP effects are effectively parameterized by
$ \kp_\lm$, $\delta_\triangle$, $\delta_\Box $,  and $ \delta_\Box^\prime $,
which change the partonic differential cross section into
\begin{eqnarray}
\label{eq:diffXShh}
&&\frac{d\hat \sigma (gg\to hh)_{\rm NP}}{d \hat t} 
\\[3pt] \nn
&=&
\frac{G_F^2 \alpha_s^2}{256 (2\pi)^3}
\left[ \; \left| \kp_\lm \frac{ 3 m_h^2}{\hat{s}-m_h^2} \big( F_\triangle + \frac{2}{3} \delta_\triangle \big) +  \big( F_\Box - \frac{2}{3} \delta_\Box \big)  \right|^2 
+\left|  G_\Box + \delta_\Box^\prime \right|^2  \right]. \hspace{0.5cm}
\end{eqnarray}

For the case \textbf{(i)}, we take the SM
except for the Higgs trilinear self-coupling.
Let us make some comments on the values of $\kp_\lm=-0.5$ and $\kp_\lm= 5.5$,
which are chosen, as simple representative numbers, 
to approximately satisfy $\sg_\np/\sg_\sm(gg\to hh)\simeq 3$.
Our calculation of the signal in what follows is at leading order.
However, 
the $K$-factor in the SM  
is not only quite large like $1.9$ at
NNLO
but also significantly varying with the transverse momentum of the Higgs boson~\cite{Grazzini:2018bsd}.
Without a reliable NLO calculation in the NP model,
tuning the value of $\kp_\lm$ to exactly get $\sg_\np/\sg_\sm|_{\rm LO}= 3$
is not of much importance.
Moreover, our main results rely on the \textit{shapes} of kinematic distributions, 
rather than the total cross section.
For the case \textbf{(ii)}, we extend the SM quark sector
by introducing new heavy VLQs.\footnote{New chiral fermions,
even in the 2HDM, 
are excluded by the Higgs precision data and the resonance searches
in the $ZZ$ and $W^+ W^-$ channel~\cite{Kang:2018jem,Belotsky:2002ym}.}
Of course, there is a possibility that both \textbf{(i)} and \textbf{(ii)} 
occur simultaneously. 
Since the combined effect
is very different
according to the relative contributions from the case \textbf{(i)} and \textbf{(ii)},
it is troublesome to quantify the result.
We do not consider the mixed case in this work.

One of the most important factors when considering the case \textbf{(ii)}
is the correlation between the di-Higgs and single-Higgs processes.
If new VLQs contribute to the di-Higgs triangle diagram, 
they cannot avoid contributing to the same single-Higgs triangle diagram.
Since the current Higgs precision data  
strongly prefer the SM-like Higgs boson,
we need to break the correlation in order to enhance the di-Higgs production rate.
We find that the key is non-SM Higgs couplings to fermions,
which demands an extension of the Higgs sector. 
In this regard, we consider a 2HDM with the VLQs in two limiting cases,
the alignment limit~\cite{Carena:2013ooa,Celis:2013rcs,Bernon:2015qea,Chang:2015goa,Das:2015mwa} (for the SM-like Higgs couplings)
and the exact wrong-sign limit~\cite{Ferreira:2014naa,Kang:2018jem} (for non-SM Higgs couplings).

\section{Brief review of the 2HDM with VLQs}
\label{sec:model}
We consider a 2HDM with VLQs, simply the VLQ-2HDM.
The SM Higgs sector is extended by introducing two complex scalar fields, 
$\Phi_1$ and $\Phi_2$.
The fermion sector also has new field components,
two additional 
$SU(2)_L$-doublet VLQs ($\mathcal{Q}_{L,R}$) and 
four $SU(2)_L$-singlet VLQs ($\mathcal{U}_{L,R}$ and $\mathcal{D}_{L,R}$):
\bea
\label{eq:phi:fields}
\hbox{two Higgs doublets:}\quad&&\Phi_i = \left( \begin{array}{c} w_i^+ \\[3pt]
\dfrac{v_i +  h_i + i \eta_i }{ \sqrt{2}}
\end{array} \right),
\quad (i=1,2),
\\[5pt] \nn
\hbox{VLQ~doublets:}\quad && \mcq_L = \left(\begin{array}{c} \mcu'_L \\ \mcd'_L \end{array} \right)
,~\mcq_R = \left(\begin{array}{c} \mcu'_R \\ \mcd'_R \end{array} \right)\,,  \\[3pt] \nn
\hbox{VLQ~singlets:}
\quad &&\mcu_L,\quad \mcu_R, \quad \mcd_L , \quad \mcd_R \,,
\eea
where $v_{1}$ and $v_2$ are the nonzero vacuum expectation values of $\Phi_{1}$
and $\Phi_{2}$ respectively,
defining $\tan\beta \equiv \tb = v_2 /v_1$. 
In what follows, we use the shorthand notation
of $s_x=\sin x$, $c_x = \cos x$ and $t_x = \tan x$ for simplicity.

In order to avoid  tree-level flavor changing neutral currents,
a discrete $Z_2$ symmetry is imposed under which 
$\Phi_1 \to \Phi_1$ and $\Phi_2 \to -\Phi_2$ \cite{Glashow:1976nt,Paschos:1976ay}.
According to the $Z_2$ parities of the fermions,
there are four types in the 2HDM: type-I, type-II, type-X and type-Y~\cite{Aoki:2009ha}.
In this work, we focus on type-II since
only it allows
the wrong-sign limit,
which will offer our key benchmark point.
The most general scalar potential with \textit{CP} invariance is written as  
\bea
\label{eq:V}
V_\Phi = && m^2 _{11} \Phi^\dagger _1 \Phi_1 + m^2 _{22} \Phi^\dagger _2 \Phi_2
-m^2 _{12} ( \Phi^\dagger _1 \Phi_2 + \hc) \nn \\
&& + \frac{1}{2}\lambda_1 (\Phi^\dagger _1 \Phi_1)^2
+ \frac{1}{2}\lambda_2 (\Phi^\dagger _2 \Phi_2 )^2
+ \lambda_3 (\Phi^\dagger _1 \Phi_1) (\Phi^\dagger _2 \Phi_2)
+ \lambda_4 (\Phi^\dagger_1 \Phi_2 ) (\Phi^\dagger _2 \Phi_1) \nn\\
&& + \frac{1}{2} \lambda_5
\left[
(\Phi^\dagger _1 \Phi_2 )^2 +  {\rm H.c.}
\right],
\eea
where $m_{11}^2$, $m_{22}^2$, and $\lambda_{1,\cdots,4}$ are real numbers
while $m^2_{12}$ and $ \lambda_5$ can be complex numbers. 
The $m_{12}^2$ term
softly breaks the $Z_2$ parity.
There are
five physical Higgs bosons: two \textit{CP}-even scalars (a light Higgs $h$ and a heavy Higgs $H$), one \textit{CP}-odd scalar $A$, 
and two charged Higgs bosons $H^\pm$~\cite{Song:2019aav}.
These mass eigenstates are 
related with the weak eigenstates in Eq.~(\ref{eq:phi:fields}) as
\begin{eqnarray}
 \left ( \begin{array}{c}
 h_1 \\
 h_2 \\
 \end{array} \right) = \mathbb{R}(\alpha)
 \left ( \begin{array}{c}
 H \\
 h \\
 \end{array} \right), \quad 
 \left ( \begin{array}{c}
 w_1^\pm \\
 w_2^\pm \\
 \end{array} \right) = \mathbb{R}(\beta)
 \left ( \begin{array}{c}
 G^\pm \\
 H^\pm \\
 \end{array} \right), \quad 
 \centering \left ( \begin{array}{c}
 \eta_1 \\
 \eta_2 \\
 \end{array} \right) = \mathbb{R}(\beta)
 \left ( \begin{array}{c}
 G^0 \\
 A \\
 \end{array} \right), 
\end{eqnarray}
where $G^\pm$ and $G^0$ are
the Goldstone bosons eaten by $W^\pm$ and $Z$ respectively. 
The rotation matrix $\mathbb{R}(\theta)$ is 
\begin{eqnarray}
\label{eq:R}
\mathbb{R}(\theta) = \left( \begin{array}{c r}
                             \cth & - \sth \\
                             \sth & \cth \\
                            \end{array} \right).
\end{eqnarray}
The SM Higgs boson is a linear combination of $h$ and $H$, given by
\bea
\label{eq:h:SM}
h_{\rm SM} = \sba h + \cba H.
\eea

\begin{table}\centering
{\renewcommand{\arraystretch}{1.4} 
	\begin{tabular}{|c||c|c|c|c|c|c|}
	\hline
limit	 &  $\kp_V$   & $\xi_V$  & $\kp_u$ & $\kp_d$ & $\kp_\lm$ & $\xi_\lm$\\  \hline\hline
alignment	  & 1 & 0 & 1 & 1& 1 & 0 \\ \hline 
EWS  	  & ~~$\frac{\tb^2-1}{\tb^2+1}$~~  & ~~$\frac{2 \tb}{\tb^2+1}$~~ & ~~1~~ & ~~$-1$~~& ~~$\frac{\tb^2-1}{\tb^2+1}$~~ & $\frac{\tb}{1+\tb^2} 
	  \left[
	   \frac{4}{3}-\frac{ 2(\mhh^2-2 M^2)}{3 \mh^2}
	  \right]$ \\ \hline
	\end{tabular}
	\caption{\label{tab:kappa} In the type-II 2HDM, the coupling modifiers
	of the \textit{CP}-even neutral Higgs bosons, 
$h$ and $H$, in the alignment limit and the exact wrong-sign (EWS) limit. 
Here $\kp_i =g_{iih}/g_{ii h_\sm}$ and
$\xi_i = g_{iiH}/g_{ii h_\sm}$ for the typical Higgs coupling $g_{iih(H)}$.
The Higgs trilinear self-coupling modifiers are named by
$\kappa_\lambda = \lambda_{hhh}/\lambda_{hhh}^{\rm{SM}}$
and 
$\xi_\lambda=\lambda_{Hhh}/\lambda_{hhh}^{\rm{SM}}$. 
Note that $M^2 \equiv m_{12}^2/(\sb\cb)$.}
	}
\end{table}

Conforming to the SM-like Higgs boson, 
we consider two limiting cases,  
the alignment limit and the exact wrong-sign (EWS) limit, 
defined by
\bea
\label{eq:def:alignment:EWS}
\hbox{alignment:} & \bt-\al=\dfrac{\pi}{2};
\\ \nn
\hbox{EWS:} & \bt+\al=\dfrac{\pi}{2}.
\eea
In these limiting cases, 
the $h$ and $H$ coupling modifiers are summarized in Table \ref{tab:kappa}.
Here 
$\kp_i =g_{iih }/g_{iih }^\sm$ 
and $\xi_i =g_{iiH}/g_{iih}^\sm$
where $g_{iih(H)}$ is a typical $h(H)$ coupling constant to gauge bosons and fermions.
For the Higgs self-coupling modifiers,
we use the convention 
$\kappa_\lambda=\lm_{hhh}/\lm_{hhh}^\sm$ and $\xi_\lambda  = \lambda_{Hhh}/\lambda_{hhh}^{\rm{SM}}$.

In the alignment limit, $h$ behaves exactly the same as $h_{\rm SM}$ 
($\kp_{i,\lm}=1$)
while the heavy Higgs $H$ is decoupled from the SM 
($\xi_{V,\lm} =0$).
Note that the resonant di-Higgs production through $gg\to H\to hh$
is absent.
In the EWS limit,
the coupling of the down-type fermion to the Higgs boson
has opposite sign to that of the up-type fermion.
Furthermore $\kp_V$ and $\kp_\lm$
deviate from the SM values
and the heavy Higgs boson $H$ is not decoupled. 
If $\tb \gg 1$, however, 
the Higgs couplings become close to the SM ones like 
$|\kp_{f,V,\lm}| \simeq 1$ and 
$\xi_\lm$ is also suppressed for large $\tb$ 
and can be further suppressed by adjusting the free parameter $m_{12}^2$. 

The Yukawa Lagrangian for the VLQs is
\bea
\label{eq:L:Yuk:0}
- {\cal L}_{\rm VLQ} &=& M_\mcq \overline{\mcq} \mcq + M_\mcu \overline{\mcu} \mcu + M_\mcd \overline{\mcd}\mcd
+ \Big[   Y_{\mcd} \bmcq \Phi_1 \mcd+
 Y_{\mcu} \bmcq \,\widetilde{\Phi}_2 \mcu + \textrm{H.c.} 
  \Big] \,,
\eea
where $\widetilde{\Phi}_i= i \tau_2 \Phi^*_i$
and we assume $Y_{\mcu (\mcd)}^L = Y_{\mcu(\mcd)}^R\equiv Y_{\mcu(\mcd)}$ for simplicity.
The VLQ mass matrices ${\mathbb M_\mcd}$ and ${\mathbb M_\mcu}$
in the basis of $(\mcd',\mcd)$ and $(\mcu',\mcu)$ are 
\begin{eqnarray}
{\mathbb M_\mcd} = \left(\begin{array}{cc} M_\mcq  & \tfrac{1}{\sqrt{2}} Y_{\mcd} v \cb  \\
\tfrac{1}{\sqrt{2}}Y_{\mcd} v \cb  & M_\mcd \end{array} \right),
\quad
{\mathbb M_\mcu} = \left(\begin{array}{cc} M_\mcq  & \tfrac{1}{\sqrt{2}} Y_{\mcu} v \sb  \\
\tfrac{1}{\sqrt{2}}Y_{\mcu} v \sb  & M_\mcu \end{array} \right).
\end{eqnarray}
The mass eigenstates are $( \mcf_{1}, \mcf_{2} )^T = \mathbb{R}(\theta_\mcf) (\mcf',\mcf)^T$
for $\mcf=\mcu,\mcd$.
The VLQ mixing angles are given by
\begin{equation}
\label{eq:mixingangle}
s_{2\theta_\mcd} = \frac{\sqrt{2}Y_\mcd v}{M_{\mcd_2}-M_{\mcd_1}}\cb\,,
\quad
s_{2\theta_\mcu} = \frac{\sqrt{2}Y_\mcu v}{M_{\mcu_2}-M_{\mcu_1}}\sb\,,
\end{equation}
where $M_{\mcu_{1,2}}$ and $M_{\mcd_{1,2}}$ are mass eigenvalues for the up-type and down-type VLQs, respectively.
We parameterize the Higgs couplings to 
the VLQ mass eigenstates by
\begin{eqnarray}
-\lg_{\rm VLQ} &\supset&
\sum_{i,j=1,2} h \Big[ y^h_{\mcd_i \mcd_j}  \overline{\mcd}_i \mcd_j
+y^h_{\mcu_i \mcu_j}  \overline{\mcu}_i \mcu_j
\Big],
\end{eqnarray}
where for $\mcf=\mcu,\mcd$ they are
\begin{eqnarray}
\label{eq:Yukh}
y^h_{\mcf_1 \mcf_1} &=& -y^h_{\mcf_2 \mcf_2} 
= - \frac{1}{\sqrt{2}}Y_\mcf \,\xi^h_\mcf \,s_{2\theta_\mcf} \,,
\\ \nn
y^h_{\mcf_1\mcf_2} &=&  y^h_{\mcf_2\mcf_1} 
= \frac{1}{\sqrt{2}} Y_\mcf \,\xi^h_\mcf \, c_{2\theta_\mcf}.
\end{eqnarray}
In type-II, $\xi^h_\mcu= \ca$ and $\xi^h_\mcd = -\sa$.

Three major constraints on the VLQ-2HDM are to be discussed.
The first one is from the Higgs precision measurements,
especially 
the loop-induced VLQ contributions to $\kp_g$: 
$\kp_\gm$ is less constrained because
the $h$-$\gamma$-$\gamma$ vertex is mainly from $W^\pm$ boson loops.
In the presence of VLQs, $\kp_g$ becomes
\begin{eqnarray}
\kappa_g = 1 + \frac{v}{ A^H_{1/2}(\tau_t)} \sum_{i=1,2}
\sum_{\mcf=\mcu,\mcd}  \frac{y^h_{\mcf_i \mcf_i}}{M_{\mcf_i}} \,
\, \,A^H_{1/2}(\tau_{\mcf_i}),
\end{eqnarray}
where $\tau_f = m_h^2/(4 m_f^2)$ and the loop function $A^H_{1/2}(\tau)$ is referred to Ref.~\cite{Djouadi:2005gi}.
The relation of $y^h_{\mcf_1 \mcf_1} = -y^h_{\mcf_2 \mcf_2}$ 
in Eq.~(\ref{eq:Yukh})
yields considerable cancelation between the contributions of $\mcf_1$ and $\mcf_2$
to $\kp_g$.
The ATLAS combined result of
$\kappa_g = 1.03^{+0.07}_{-0.06}$ 
\cite{Aad:2019mbh} is satisfied in most of the parameter space.

The second constraint is from the direct searches for VLQs
at the LHC
by the ATLAS~\cite{Aad:2015mba,Aad:2015kqa,Aad:2015voa,Aad:2016qpo,Aad:2016shx,Aaboud:2017qpr,Aaboud:2017zfn,
Aaboud:2018xuw,Aaboud:2018uek,Aaboud:2018saj,Aaboud:2018wxv,Aaboud:2018pii} and
CMS~\cite{Chatrchyan:2013wfa,Khachatryan:2015gza,Khachatryan:2015oba,Khachatryan:2016vph,
Sirunyan:2016ipo,Sirunyan:2017ezy,Sirunyan:2017tfc,Sirunyan:2017usq,Sirunyan:2017ynj,Spiezia:2017ueo} collaborations.
The lower mass bounds on the VLQs depend 
sensitively on the decay channels of VLQs:
if
they decay only into the third generation quarks, 
the bounds are strong such that
$M_\mcu > 1.31\tev$  and $M_\mcd > 1.03\tev$~\cite{Aaboud:2018pii}.
The  bounds are relaxed into $M_\mcq > 690\gev$
if the VLQ 
decays into a light quark $q$~\cite{Aad:2015tba}. 
If $H^\pm q$ mode is open, the VLQ mass bound will further be  weakened.
In what follows,
therefore,
we take the case of $M_{\mcf_1} \gsim 600\gev$.

Finally, we consider the strongest constraint on the VLQ-2HDM
from the electroweak precision data,
the Peskin-Takeuchi oblique parameters $S$, $T$, and $U$~\cite{Peskin:1991sw,Barbieri:2004qk}.
Based on more general parametrization 
in terms of $\hat S$, $\hat T$, $W$, and $Y$~\cite{Barbieri:2004qk},
we found in the previous work~\cite{Song:2019aav} 
that the most sensitive oblique parameter $\hat{T}$ vanishes in the following ansatz: 
\bea
\label{eq:ansatz}
\hbox{zero-}\hat{T}\hbox{ ansatz: }
M_{\mcu_1}=M_{\mcd_1}  \equiv M_1 , \quad 
M_{\mcu_2}=M_{\mcd_2}  \equiv M_2  , \quad 
 \theta_\mcu =\theta_\mcd  \equiv \theta  .
\eea
In this ansatz,
the up-type and down-type VLQ Yukawa couplings are related as
\bea
Y_\mcu s_\beta = Y_\mcd c_\beta = \frac{s_{2\theta}\Delta M}{\sqrt{2} v},
\eea
where $\Dt M =M_{2} - M_{1}$.
Then the Higgs Yukawa couplings to the VLQs in Eq.~(\ref{eq:Yukh})
take the simple forms of
\bea
\label{eq:y:zeroT}
\hbox{alignment:} && y^h_{\mcu_1 \mcu_1}= y^h_{\mcd_1 \mcd_1} 
                           = -y^h_{\mcu_2 \mcu_2}= -y^h_{\mcd_2 \mcd_2} 
=-\frac{\Dt M}{2 v} s_{2 \theta}^2;
\\[3pt] \nn
&& y^h_{\mcu_1\mcu_2} = y^h_{\mcd_1\mcd_2} 
     =y^h_{\mcu_2\mcu_1} = y^h_{\mcd_2\mcd_1} 
= \frac{\Dt M}{2 v}c_{2 \theta} s_{2 \theta};
\\[5pt] \nn
\hbox{EWS:} && y^h_{\mcu_1 \mcu_1}= -y^h_{\mcd_1 \mcd_1} 
                      =-y^h_{\mcu_2 \mcu_2}= y^h_{\mcd_2 \mcd_2}
=-\frac{\Dt M}{2 v} s_{2 \theta}^2;
\\[3pt] \nn
&& y^h_{\mcu_1\mcu_2} = - y^h_{\mcd_1\mcd_2} 
    = y^h_{\mcu_2\mcu_1} = - y^h_{\mcd_2\mcd_1} 
= \frac{\Dt M}{2 v}c_{2 \theta} s_{2 \theta}.
\eea
In the EWS limit, the down-type VLQ Higgs coupling is equal and opposite
to the up-type one, while in the alignment limit they are the same.
This feature will determine the correlation between the VLQ contributions
to the single-Higgs and di-Higgs production rates.

\section{Characteristics of the non-resonant NP effects on the di-Higgs process}
\label{sec:characteristics}


In this section, we study the phenomenological characteristics of 
different NP effects on the non-resonant di-Higgs process.
First we need to find a reasonable benchmark point in the VLQ-2HDM,
satisfying $\sg_\np/\sg_\sm (gg \to hh) \simeq 3$
and $\sg_\np/\sg_\sm (gg \to h) \simeq 1$ simultaneously.
Equation (\ref{eq:diffXShh}),
$\sg_\np/\sg_\sm(gg \to hh)$
in terms of $ \delta_\triangle$, $ \delta_\Box$, and $ \delta_\Box^\prime$,
will help the exploration.
In the alignment limit which guarantees $\kp_\lm=1$,
the ratio at the 14 TeV LHC is 
\begin{eqnarray}
\label{eq:XSdev:SM}
\left. \frac{\sigma (gg\to hh)_{\rm NP}}{ \sigma (gg\to hh)_{\rm SM}} 
\right|_{\kp_\lm=1}
 &=&1 - 0.37 \,\delta_\triangle + 0.92 \,\delta_\Box  - 0.28 \,\delta_\Box^\prime 
 \\ \nn &&
 + \, 0.13 \,\delta_\triangle^2 + 1.57 \,\delta_\Box^2 
 + 3.54 \,\delta_\Box^{\prime 2} - 0.62 \,\delta_\triangle \delta_\Box\,,
\end{eqnarray}
where \textsc{Nnpdf30} parton distribution function set is used.
In the EWS limit, $\kp_\lm$ is slightly deviated from one:
for $\tb=5$, $\kp_\lm \simeq 0.92$ and the ratio is 
\begin{eqnarray}
\label{eq:XSdevEWS}
\left. \frac{\sigma (gg\to hh)_{\rm NP}}{ \sigma (gg\to hh)_{\rm SM}} 
\right|_{\kp_\lm=0.92}
 &=& 1.06 - 0.36 \,\delta_\triangle + 0.98 \,\delta_\Box  - 0.28 \,\delta_\Box^\prime 
 \\ \nn &&
 + \,0.11 \,\delta_\triangle^2 + 1.57 \,\delta_\Box^2 
 + 3.54 \,\delta_\Box^{\prime 2} - 0.57 \,\delta_\triangle \delta_\Box\,.
 \end{eqnarray}

We analytically calculate the new form factors 
with finite VLQ masses,
which are almost consistent with the formulae in Ref.~\cite{Asakawa:2010xj}.\footnote{We found several typos in Ref.~\cite{Asakawa:2010xj}.
In  Eq.~(B12), there are three typos:
(i) the overall sign in the right-hand-side should be $(+)$;
(ii)  
``$-4(D_{27[t,t,t,T]}^{(1,2,3)}+\cdots$''   should be 
``$-8(D_{27[t,t,t,T]}^{(1,2,3)}+\cdots$'';
(iii) 
 ``$\cdots -C_{[t,t,T]}^{(3,4)}) \}$''  should be 
 ``$\cdots -\frac{1}{2} C_{[t,t,T]}^{(3,4)}) \}$''.
In Eq. (B13), we should replace 
``$-16\left(\frac{\epsilon_t}{\sqrt{2}} \right )\left(\frac{\epsilon_T}{\sqrt{2}} \right) m_t m_T( \cdots$''   by  ``$-32\left(\frac{\epsilon_t}{\sqrt{2}} \right )\left(\frac{\epsilon_T}{\sqrt{2}} \right) m_t m_T( \cdots$''.
}
In order to double-check, 
we derived the asymptotic behaviors of the new form factors in the LET,
and found them completely consistent with those in Ref.~\cite{Dawson:2012mk}.
For $M_{\mcf} \gg 2 m_h$, 
the new form factors are
\begin{eqnarray}
\label{eq:largemassdeltas}
\delta_\triangle &\simeq&  \sum_{i=1,2}
\left[
\frac{v}{M_{\mcu_i}}y_{\mcu_i \mcu_i}^h
+
\frac{v}{M_{\mcd_i}}y_{\mcd_i \mcd_i}^h
\right]\,, \\ \nn
\delta_\Box &\simeq &
\sum_{i=1,2}
\left[
\frac{v^2}{M_{\mcu_i}^2} \lf y_{\mcu_i \mcu_i}^h \ri^2
+
\frac{v^2}{M_{\mcd_i}^2} \lf y_{\mcd_i \mcd_i}^h \ri^2
\right]
+ \sum_{\mcf=\mcu,\mcd} \frac{2v^2}{M_{\mcf_1}M_{\mcf_2}}\lf y_{\mcf_1 \mcf_2}^h \ri^2,
\\ \nn
\delta_\Box^\prime &\simeq& 0
\,.
\end{eqnarray}

Adopting the zero-$\hat T$ ansatz in Eq.~(\ref{eq:ansatz}), 
where $y^h_{\mcu_i \mcu_i} = y^h_{\mcd_i \mcd_i}$ in the alignment limit
while
$y^h_{\mcu_i \mcu_i} = - y^h_{\mcd_i \mcd_i}$ in the EWS limit,
the NP form factors are further simplified as
\bea
\label{eq:T0deltas}
\delta_\triangle^{\textrm{zero-}\hat T} &\simeq&
\left\{
\barr{ll}
-\frac{(\Delta M)^2 }{M_1 M_2}s_{2\theta}^2 & ~~~\hbox{(alignment);} \\[3pt]
0 & ~~~\hbox{(EWS);}
\earr
\right.
\\ [7pt] \label{eq:dt:box}
\delta_\Box^{\textrm{zero-}\hat T} &\simeq&
\barr{ll}
\frac{(\Delta M)^2 }{M_1 M_2}s_{2\theta}^2 
+ \frac{1}{2}\frac{(\Delta M)^4 }{M_1^2 M_2^2} s_{2\theta}^4 
& ~~~\hbox{(alignment \& EWS).} 
\earr
\eea
As shown in Eqs.~(\ref{eq:XSdev:SM}) and (\ref{eq:dt:box}), 
the contributions from the box diagrams in both limits
are positive to the SM contribution.
Moreover, $\delta_\Box$ is proportional to the 
quadratic or quartic terms of the VLQ mass difference $\Dt M$:
we need sizable $\Dt M$ to enhance 
the di-Higgs production rate.
In the alignment limit,
large $\Dt M$ also increases $\dt_\triangle$
and thus the contribution to the single-Higgs production rate.
In the EWS limit, however, 
$\dt_\triangle$ is negligible because of the relation of 
$y^h_{\mcu_i \mcu_i}=-y^h_{\mcd_i \mcd_i}$: see Eq.~(\ref{eq:y:zeroT}).

\begin{figure}[h] \centering
\includegraphics[width=0.48\textwidth]{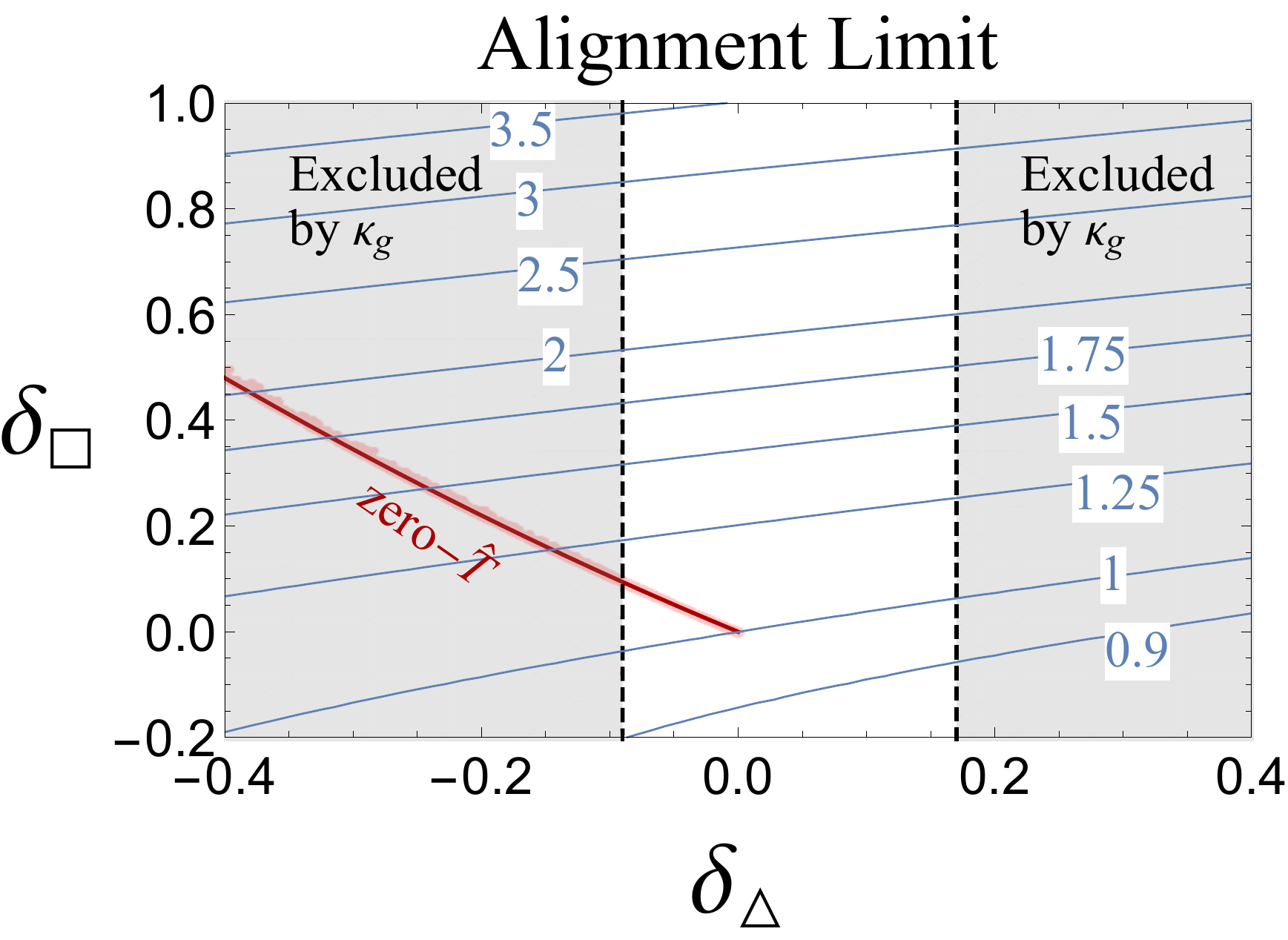}~
\includegraphics[width=0.48\textwidth]{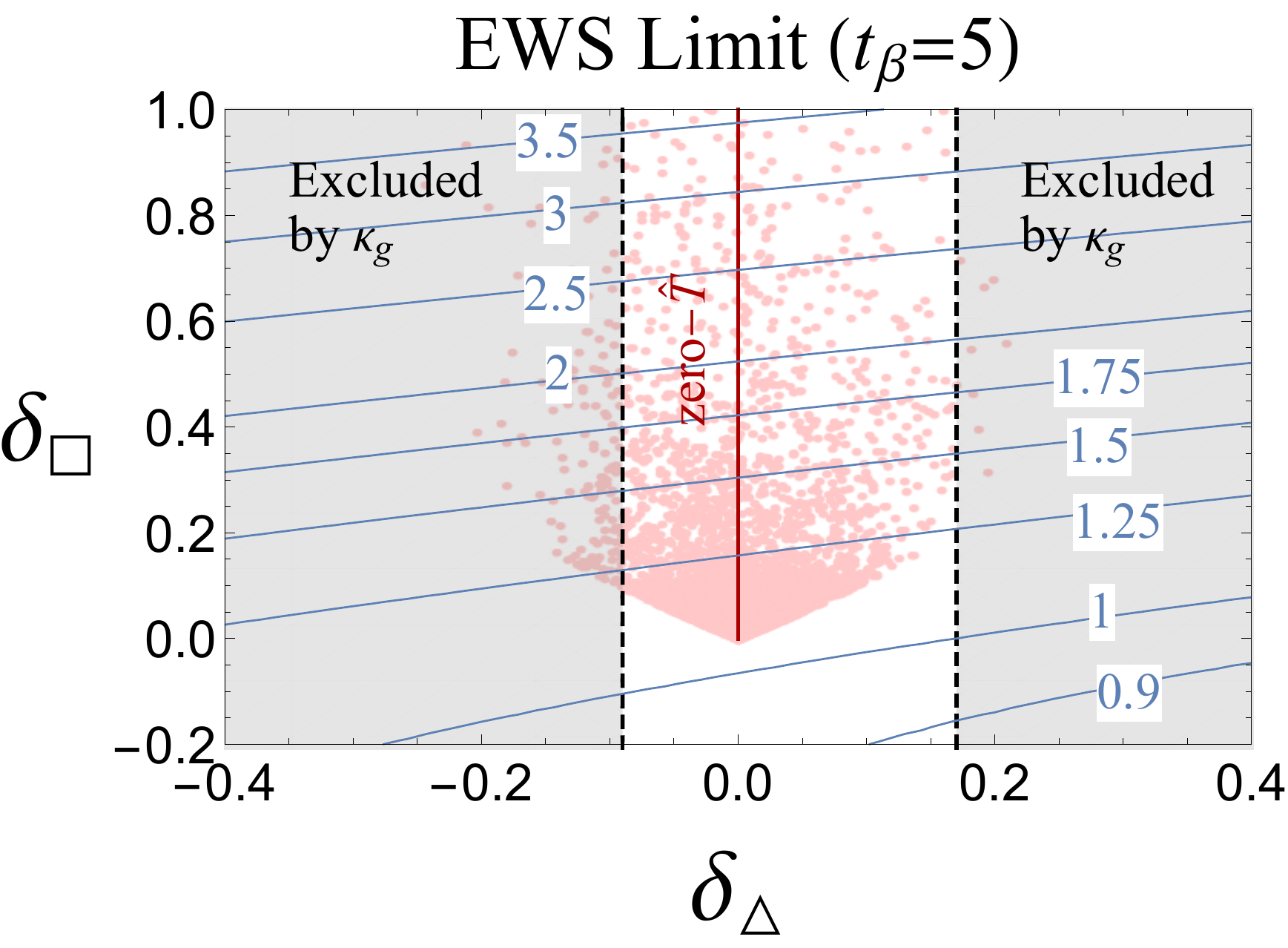}
\caption{\baselineskip 3.5ex
The VLQ-2HDM prediction of the di-Higgs production rate 
and various constraints 
on $(\delta_\triangle,\delta_\Box)$
in the alignment (left panel) and EWS (right panel) limits. 
We set $\tb=5$.  
The blue contours denote ${\sigma _{\rm NP}}/{ \sigma _{\rm SM}}(gg\to hh)$ 
by assuming  $\dt_\Box^\prime = 0$.
The red scatter dots are allowed by the electroweak oblique parameters at $2\sg$,
the direct LHC search bounds on the VLQ masses, and the perturbativity
of the Yukawa coupling.
The red lines are the results of the zero-$\hat{T}$ ansatz.
The grey regions are excluded by the current measurement 
on the Higgs coupling modifier $\kp_g$ at $2\sg$.
}
\label{fig:scatterdeltas}
\end{figure} 

More correlations between the di-Higgs production rate and other constraints
are summarized in Fig.~\ref{fig:scatterdeltas}.
Over the parameter space $(\delta_\triangle,\delta_\Box)$, 
we present the contours of ${\sigma_{\rm NP}}/{ \sigma_{\rm SM}(gg\to hh)}$
(blue lines)
in the VLQ-2HDM for the alignment limit (left panel) and EWS limit (right panel)
with $\dt_\Box^\prime = 0$ and $\tb=5$. 
As can be seen from the slopes of the contours,
${\sigma_{\rm NP}}/{ \sigma_{\rm SM}}$ depends 
more sensitively on $\dt_\Box$ than $\dt_\triangle$.
This is attributed to the larger coefficients of  $\dt_\Box$ and $\dt_\Box^2$ 
than those of $\dt_\triangle$ and $\delta^2_\Delta$  in
Eqs.~(\ref{eq:XSdev:SM}) and (\ref{eq:XSdevEWS}).
The constraints from the electroweak oblique parameter $\hat{T}$
along with the LHC direct searches for the VLQ
and the perturbativity of Yukawa couplings
are shown by the scatter plots.
The red dots are allowed by the oblique parameter $\hat{T}$ at $2\sg$~\cite{Tanabashi:2018oca},
through scanning the parameters over the following range:
\begin{eqnarray}
M_{\mcu_{1,2}}, M_{\mcd_{1,2}} > 600\gev, \quad
{\bar Y}_\mcu (\equiv Y_\mcu s_\beta),~ {\bar Y}_\mcd (\equiv Y_\mcd c_\beta) < 4\pi\,.
\end{eqnarray}
Additionally, we present the results of the zero-$\hat{T}$ ansatz by red lines.
Finally we show the $2\sg$ exclusion region (grey areas)
by the current Higgs precision data 
of $\kp_g = 1.03^{+0.07}_{-0.06}$~\cite{Aad:2019mbh}

The alignment and EWS limits exhibit very different behaviors.
In the alignment limit,
the result of
the zero-$\hat{T}$ ansatz (red line)
shows a strong correlation of $\delta_\Box  \approx - \delta_\triangle$. 
In addition,
all of the red dots
are closely gathered around the zero-$\hat{T}$ ansatz line.
A large $\delta_\Box$ inevitably leads to a large $\delta_\triangle$,
which is severely limited by the single-Higgs production rate such as $\left| \dt_\triangle \right| \lsim 0.1$.
In the alignment limit, therefore,
the current LHC Higgs precision data 
permit at most $20\%$ increase in the di-Higgs production rate.
In the EWS limit,
the zero-$\hat{T}$ ansatz (red line) guarantees $\dt_\triangle\simeq 0$
so that the constraint from $\kp_g$ becomes negligible.
Relaxing the $\hat{T}$ constraint within $2\sg$ (red dots)
allows much wider spread of the allowed parameter points in $(\delta_\triangle,\delta_\Box)$,
quite far from the red line.

On account of the overall features in Fig.~\ref{fig:scatterdeltas},
we take the following benchmark point in the EWS limit 
for our basic assumption $\sg_\np/\sg_\sm (gg\to hh)\simeq 3$:
\bea
\label{eq:benchmark}
\hbox{benchmark: } && \beta+\al=\frac{\pi}{2},
\quad \tb=5, 
\\ \nn && M_1 = 600\gev,\quad \Dt M = 900\gev,\quad \theta=0.6.
\eea
We find that the contributions from $\mcu_2$ and $\mcd_2$
are negligible, below $\sim 1\%$.


\begin{figure}[h] \centering
\includegraphics[height=190pt]{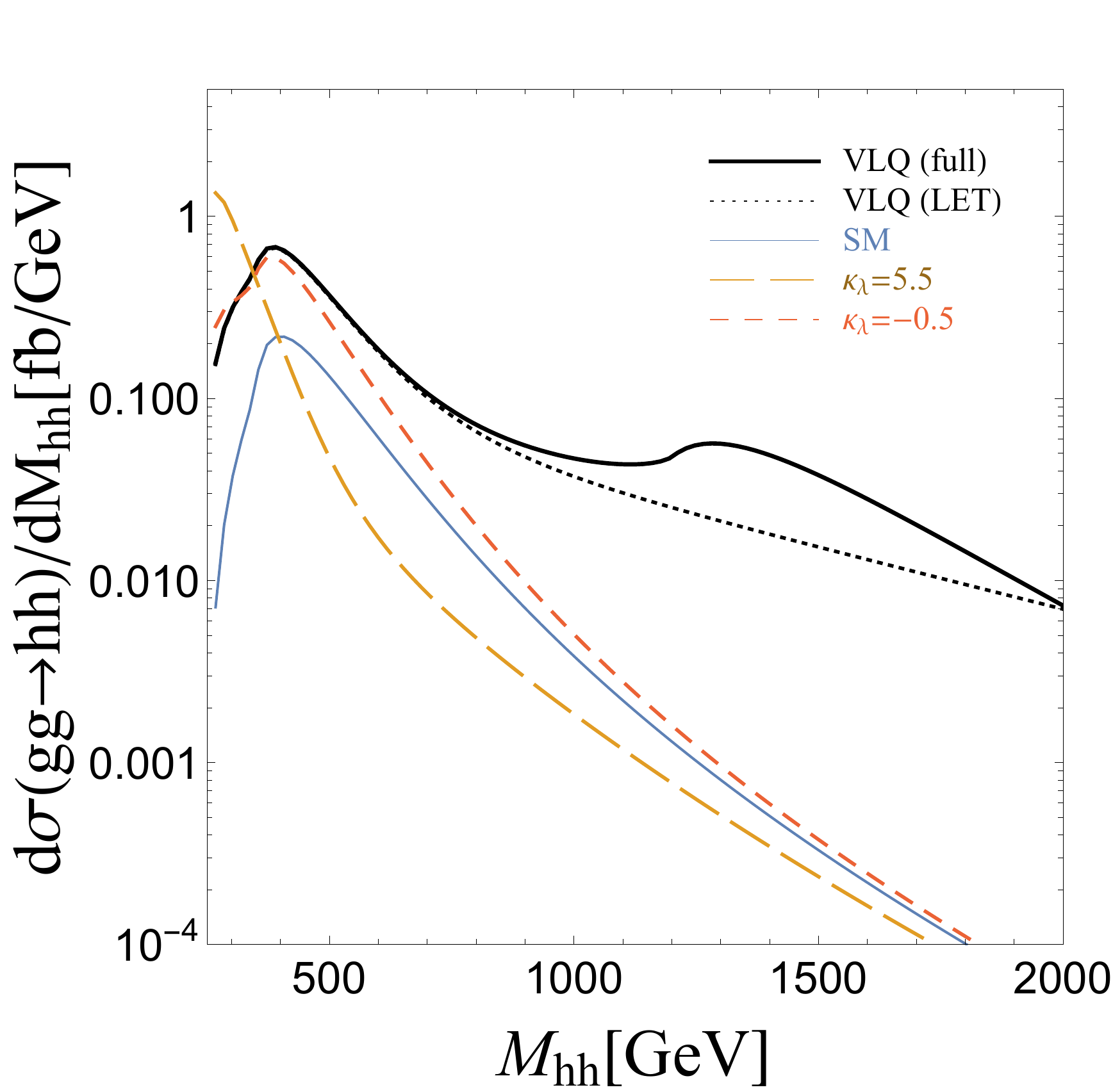}
\includegraphics[height=190pt]{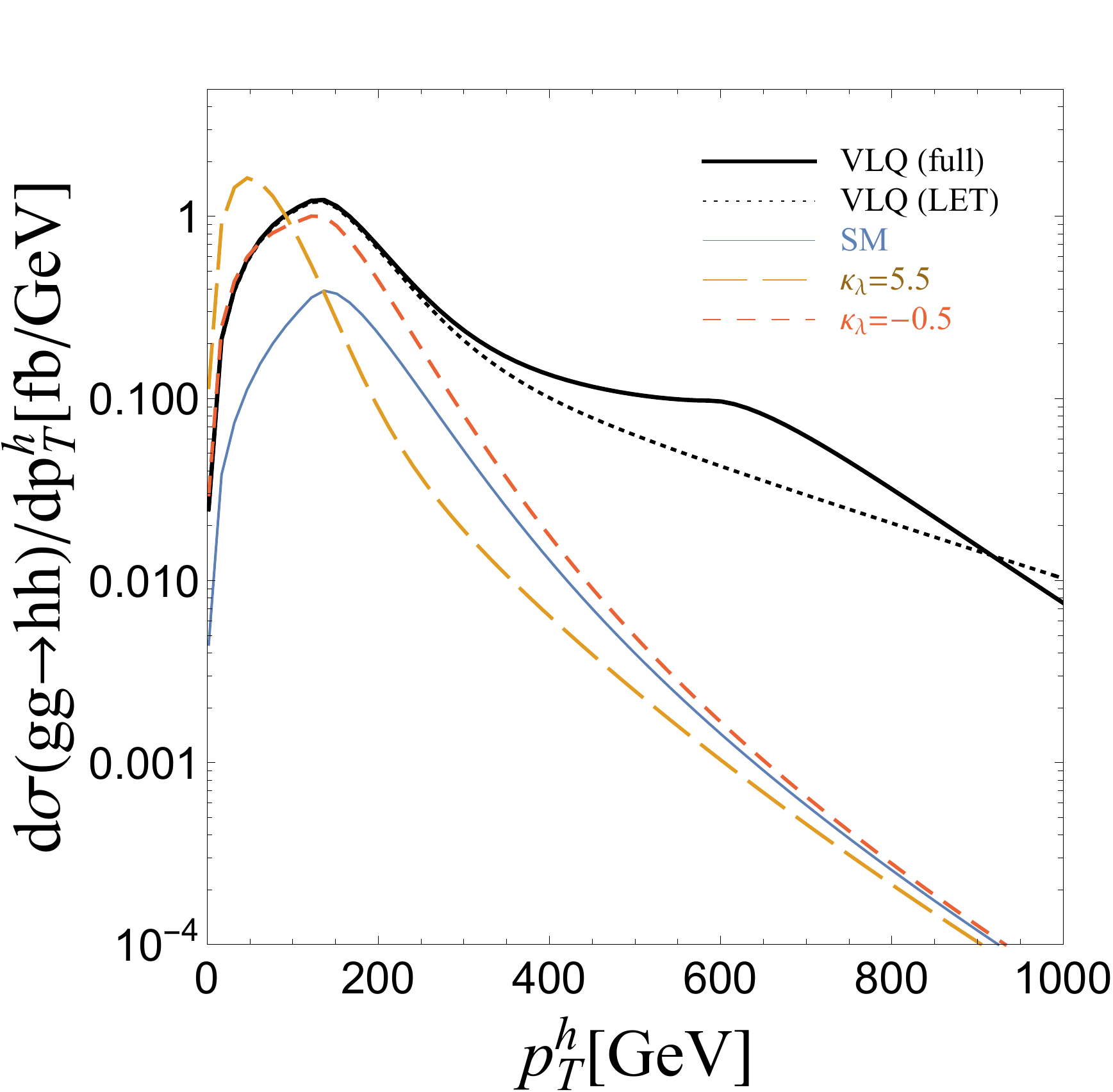}
\caption{\baselineskip 3.5ex 
The distributions of the invariant mass of of the Higgs-boson pair (left panel) and 
those of the transverse momentum of of one of the Higgs bosons (right panel)
for the parton level $gg\to hh$ process at the 14 TeV LHC.
We consider the VLQ-2HDM with full calculations of the form factors (black solid line),
the VLQ-2HDM with the low energy theorem approximation (black dotted line),
the SM (blue solid line),
the SM with $\kp_\lm=5.5$ (brown long dashed line)
and $\kp_\lm=-0.5$ (orange dashed line).
For the VLQ-2HDM, we use the benchmark point in Eq.~(\ref{eq:benchmark}).
}
\label{fig:diffXS}
\end{figure}

Now 
we show the $M_{hh}$ (left panel) and $p_{T}^{h}$ (right panel)
distributions of the di-Higgs process
at the 14 TeV LHC in Fig.~\ref{fig:diffXS}.
We consider the VLQ-2HDM with full calculations of the form factors (black solid line),
the VLQ-2HDM with the LET approximation (black dotted line),
the SM with $\kp_\lm=5.5$ (yellow long dashed line)
and the SM with $\kp_\lm=-0.5$ (orange dashed line).
As a reference, we also present the SM results (blue solid line).
All of the results are at the parton level 
with the NNLO K-factor $K=1.85$~\cite{Borowka:2016ehy,Borowka:2016ypz,Grazzini:2018bsd,Baglio:2018lrj,Baglio:2020ini}.
Obviously,
the $M_{hh}$ and $p_T^h$ distributions 
in different NP models show meaningful differences.
For $\kp_\lm=-0.5$, 
both $M_{hh}$ and $p_T^h$ distributions 
slightly shift toward lower region, compared with those in the SM.
If $\kp_\lm=5.5$,
the shift is also to the left but much more significant such that
the peak positions in both distributions move 
about $100\gev$.
In the VLQ-2HDM,
both differential cross sections decrease much slowly as $M_{hh}$ or $p_T^h$ increases. 
It is because the box diagrams from VLQs, 
which mainly enhance the di-Higgs process,
do not have the $1/{\hat s}$ suppression at the amplitude level as in Eq.~(\ref{eq:diffXShh}).
Most of all,
we do see 
the threshold effects appear as the bump structures 
at the positions of
$M_{hh} \simeq 2 M_1$ and $p_T^h \simeq M_1$. 
Actually, the bumps lift both distributions up in the high-mass and high-$p_T$ regions.
Note that if we use the approximated form factors for the VLQ-2HDM (black dotted lines),
the bump structures disappear.

In order to show the differences quantitatively,
we calculate the ratio of the di-Higgs production cross section
after $p_T^h > 300\gev$ cut to their corresponding total cross section:
\bea
\label{eq:ratio:sigma}
\frac{\sg(gg\to hh; p_T^h > 300\gev)}{\sg_\tot(gg\to hh)}
=\left\{
\barr{rl}
6.1\%, & \quad\hbox{ (SM)} \\
14.5\%, & \quad\hbox{ (VLQ-2HDM)} \\
3.2\%, & \quad\hbox{ }(\kp_\lm=-0.5) \\
1.2\%. & \quad\hbox{ } (\kp_\lm=5.5)\\
\earr
\right.
\eea
We caution the readers that the above results are based on the parton level calculation,
so the results may vary according to the final state in full collider simulation.
The results in Eq.~(\ref{eq:ratio:sigma}) clearly show that
the high $p_T^h$ cut saves considerable amount of the VLQ-2HDM events.
This is a smoking-gun signature of the VLQ contributions to the di-Higgs process.

\section{Simulations, event selections, and analysis at the 14 TeV HL-LHC}
\label{sec:simulation}

In the previous section,
we showed that the effects of VLQs on the di-Higgs process
could be distinguished from those of non-SM Higgs trilinear self-coupling
by the correlated threshold structures
in the $M_{hh}$ and $p_T^h$ distributions.
However, the di-Higgs channel has a very small production cross section, 
raising the concern whether the characteristic feature
disappears in actual experiments.
In this section, we present the full collider simulation of the signals in two final states, 
$hh\to \bb\bb$ and $hh \to \bb\rr$.
The $4b$ final state has the advantage of the largest branching ratio of 
$\br(hh\to 4b) \sim 1/3$,
which has the second-highest sensitivity next to the $\bb\ttau$ final state~\cite{Aad:2019uzh}.
Another important final state 
is $\bb\rr$, which benefits from clean signal extraction because of a good 
di-photon invariant mass resolution,
despite much smaller branching ratio $\br(hh\to \bb\rr) \simeq 2.6 \times 10^{-3} $. 
Although we do not make a full signal-to-background selection analysis here,
the correlations 
among the key observables of the di-Higgs process may help in designing new search strategies 
for the possibility of having VLQs. 

The signal events are generated at leading order by using  
\textsc{Madgraph5\_aMC@NLO} \cite{Alwall:2011uj, Alwall:2014hca}
in the SM, the VLQ-2HDM, the SM with $\kp_\lm=-0.5$, and $\kp_\lm=5.5$.
The VLQ-2HDM model file in the \textsc{Ufo} format
is obtained from modifying an existing 2HDM model file 
by adding the new contributions of VLQs.
We thoroughly checked the \textsc{Ufo} file
by comparing various results with the analytic calculations at parton level.
All of the VLQ-2HDM results in this section are 
based on the benchmark point in Eq.~(\ref{eq:benchmark}).
We have chosen the renormalization and factorization scales to be 
twice the mass of the SM Higgs boson. 
We employ the \textsc{Nnpdf30\_lo} PDF set with $\alpha_s(M_Z)=0.118$~\cite{Ball:2014uwa}. 
The generated events are passed to \textsc{Pythia8}~\cite{Sjostrand:2014zea} 
for parton showering and hadronization, 
without multiple-parton interactions. 
We use \textsc{Delphes} as a fast detector simulation~\cite{deFavereau:2013fsa}
with the ATLAS template. 
Jets are clustered 
using the anti-$k_T$ algorithm~\cite{Cacciari:2008gp} with a jet radius of $R=0.4$ 
as implemented in \textsc{FastJets}~\cite{Cacciari:2011ma}.

\subsection{$\bb\bb$ final state}

For the $\bb\bb$ final state,
we follow the ATLAS analysis strategy~\cite{Aad:2015uka}. 
We start the event selection by requiring the presence of at least four $b$-jets with $p_T^b > 40\gev$ and $|\eta^b| < 2.5$. The four leading $b$-jets,
ordered by the transverse momentum of each $b$ jet,
are used to form two separate dijets: 
two $b$-jets with the angular distance ($\Delta R = \sqrt{\Dt \eta^2 + \Dt \phi^2}$) 
smaller than $1.5$
are identified as one dijet system.
This selection step reduces the number of events in the SM
by a factor of about $2$.


\begin{figure}[!t]
    \centering
    \includegraphics[width=0.49\textwidth]{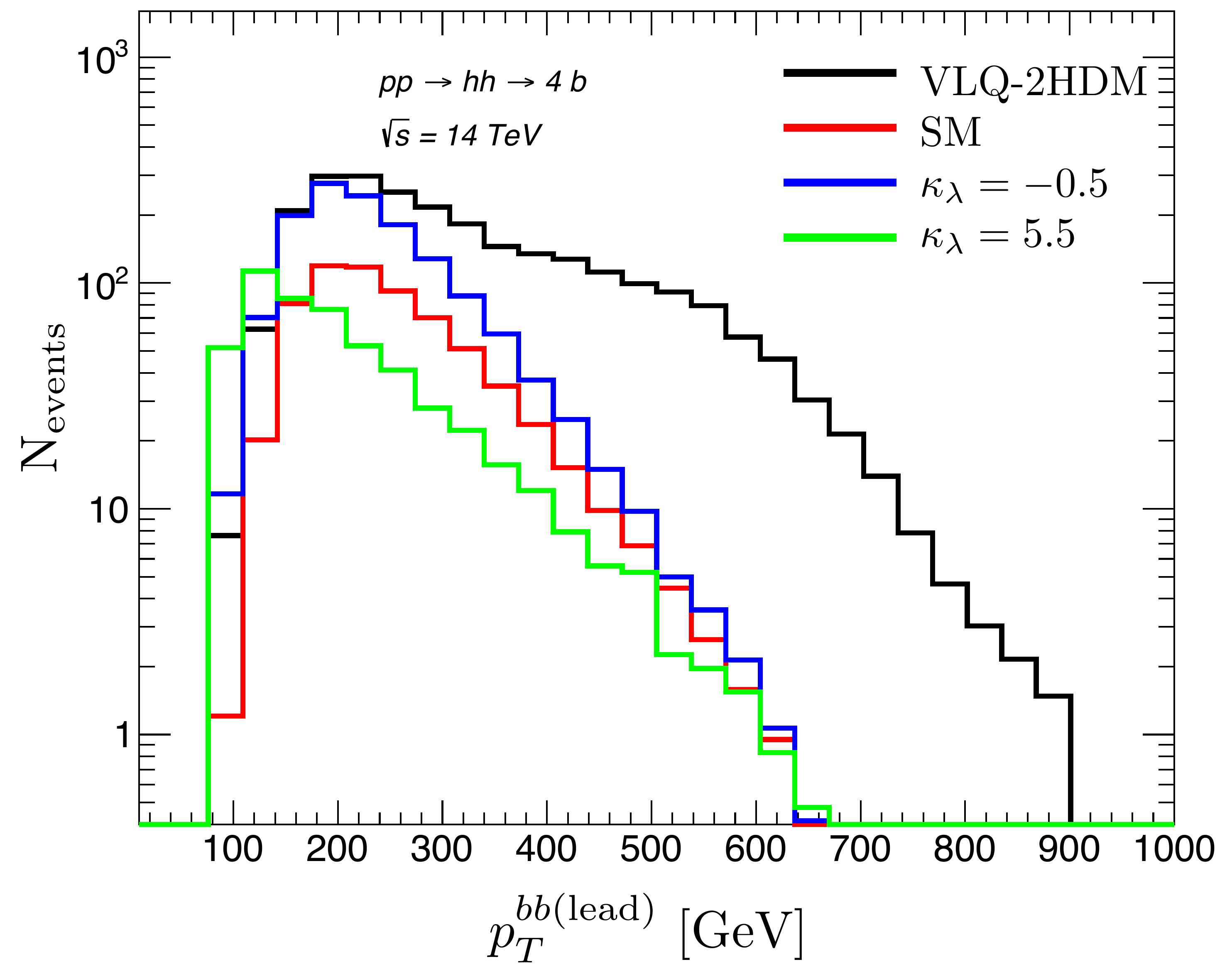}
    \includegraphics[width=0.49\textwidth]{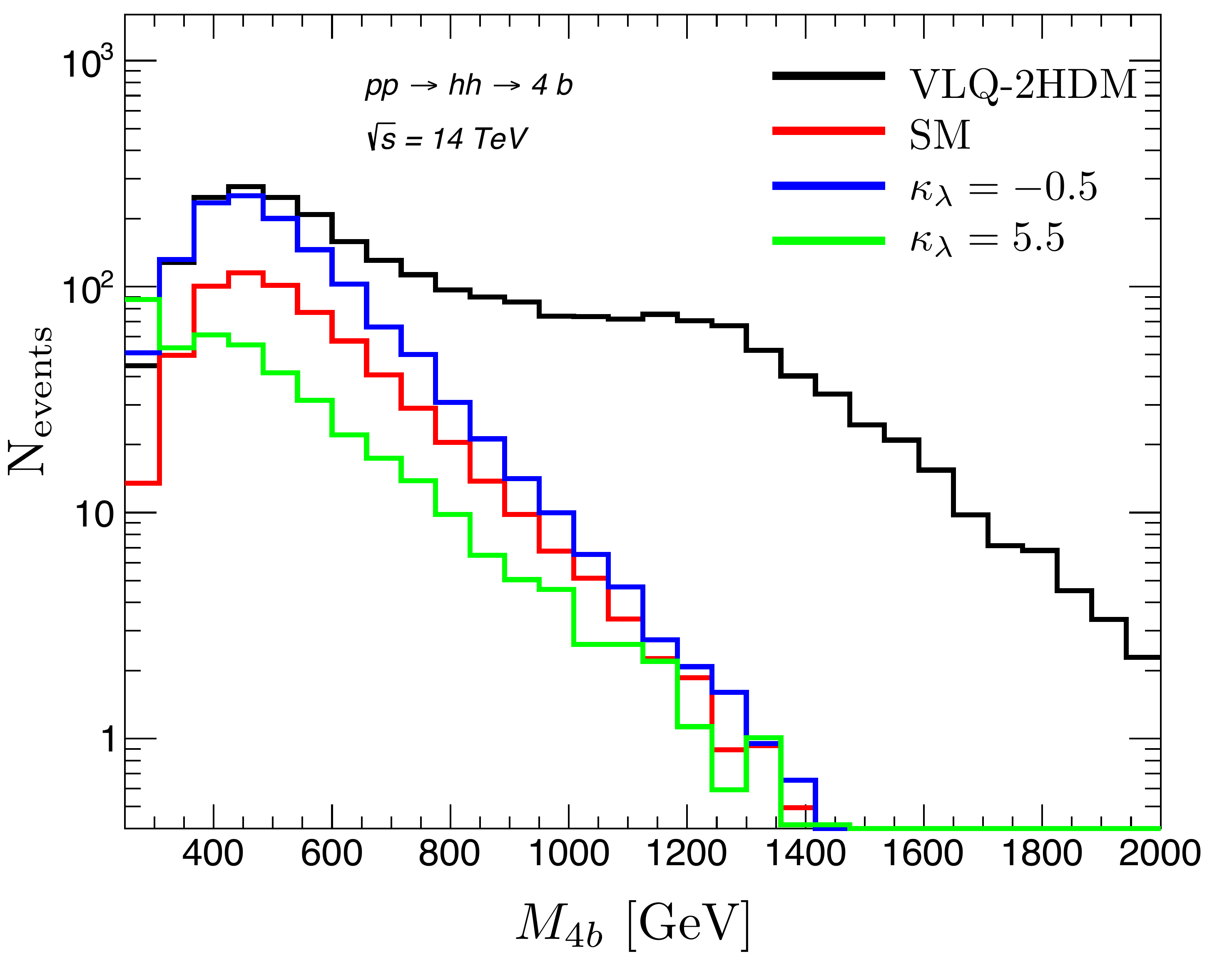}
    \caption{The expected number of events,
    after the basic selection, as a function 
    of the transverse momentum of the leading dijet
    (left panel) 
    and the invariant mass of four leading $b$-jets (right panel) 
    for $gg \to hh \to \bb\bb$ at the 14 TeV
    LHC with the total integrated luminosity $\mathcal{L} = 3000$ fb$^{-1}$. 
    The distributions are for the VLQ-2HDM (black line), the SM  (red), 
    the SM with $\kp_\lm=-0.5$ (blue), and $\kp_\lm=  5.5$ (green). 
    }
    \label{fig:m4j:pT2j1}
\end{figure}

In Fig. \ref{fig:m4j:pT2j1}, 
we show the distributions of the transverse momentum 
of the leading dijet $p_T^{bb(\textrm{lead})}$
(left panel) 
and the invariant mass of the $4b$ system 
for $gg \to hh \to 4b$
in the VLQ-2HDM (black), the SM  (red), 
the SM  with $\kp_\lm=-0.5$ (blue), and $\kp_\lm=  5.5$ (green). 
We first remark that in the $\kp_\lm = 5.5$ case, 
the total number of events
(originally corresponding to $\sg_\np/\sg_\sm(gg\to hh) \simeq 3$)
is considerably reduced and
the peaks of both distributions are shifted toward 
low values.
This is because some $b$-jets in the event 
are too soft to pass the first selection $p_T^b > 40\gev$~\cite{Chang:2018uwu}.
An encouraging observation is that the threshold effects of the VLQs are visible at the reconstruction level.
We can clearly see two bump-like structures in both $p_T^{bb(\textrm{lead})}$ and $M_{4b}$ distributions,
peaked at $p_T^{bb(\textrm{lead})} \sim M_1$ and $M_{4b} \sim 2 M_1$,
with a minor smearing effect due to the detector angularity. 
Since the two peak positions are closely related, a study of the correlation 
between the two observables will be extremely useful to probe new VLQs in the di-Higgs process.

\begin{figure}[t!]
    \centering
    \includegraphics[height=180pt]{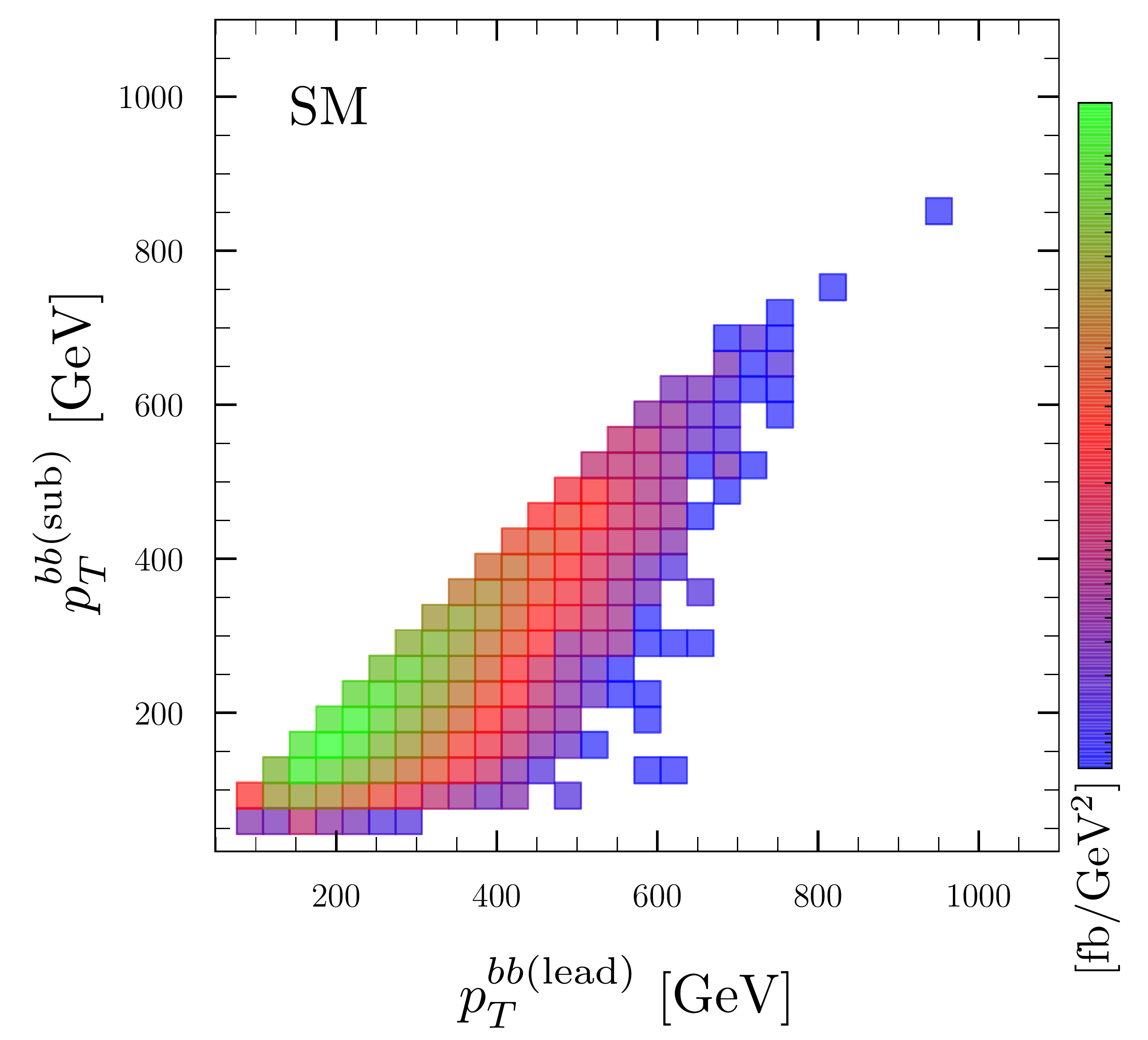}
    \includegraphics[height=180pt]{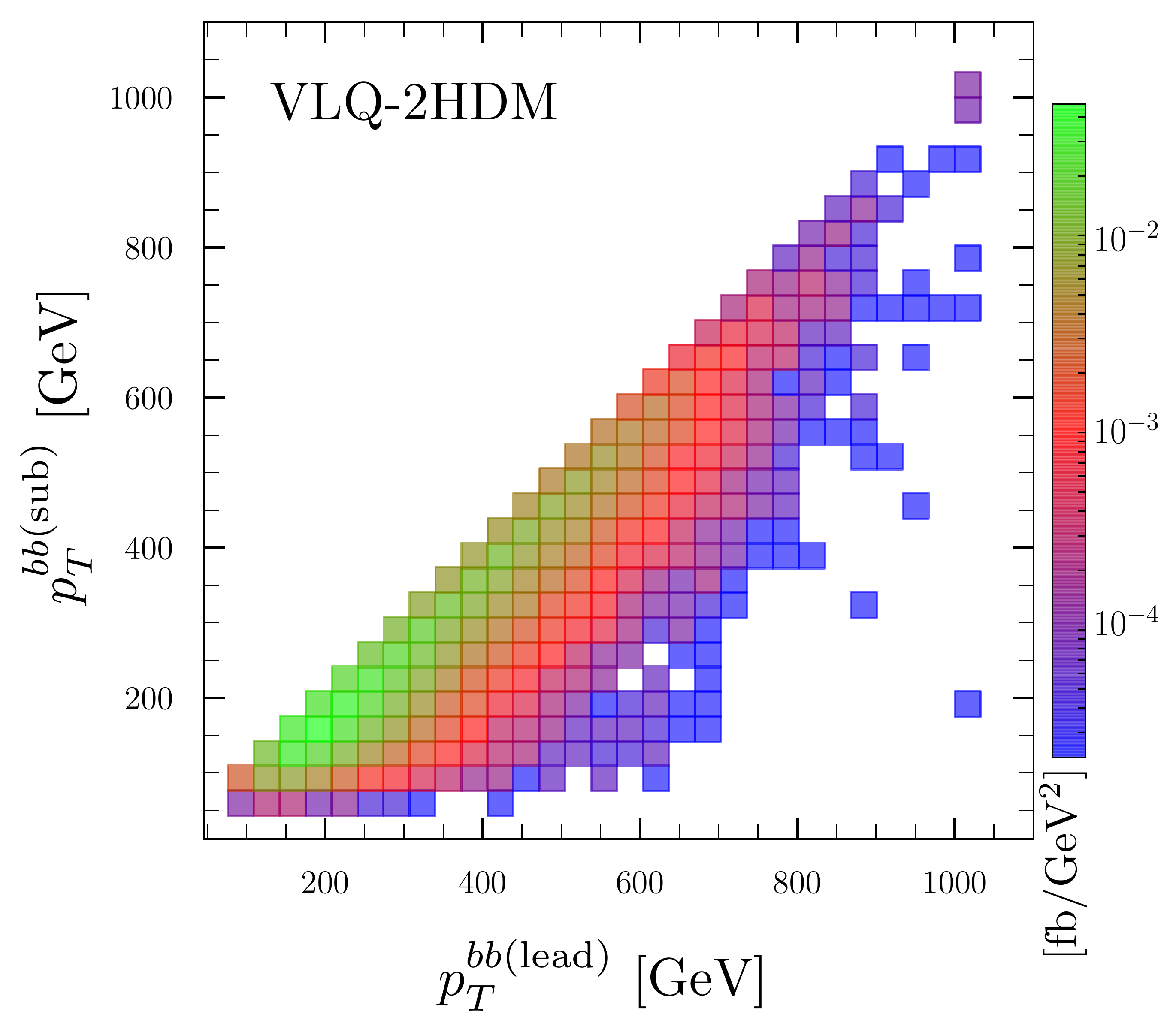}\\[12pt]
    \includegraphics[height=180pt]{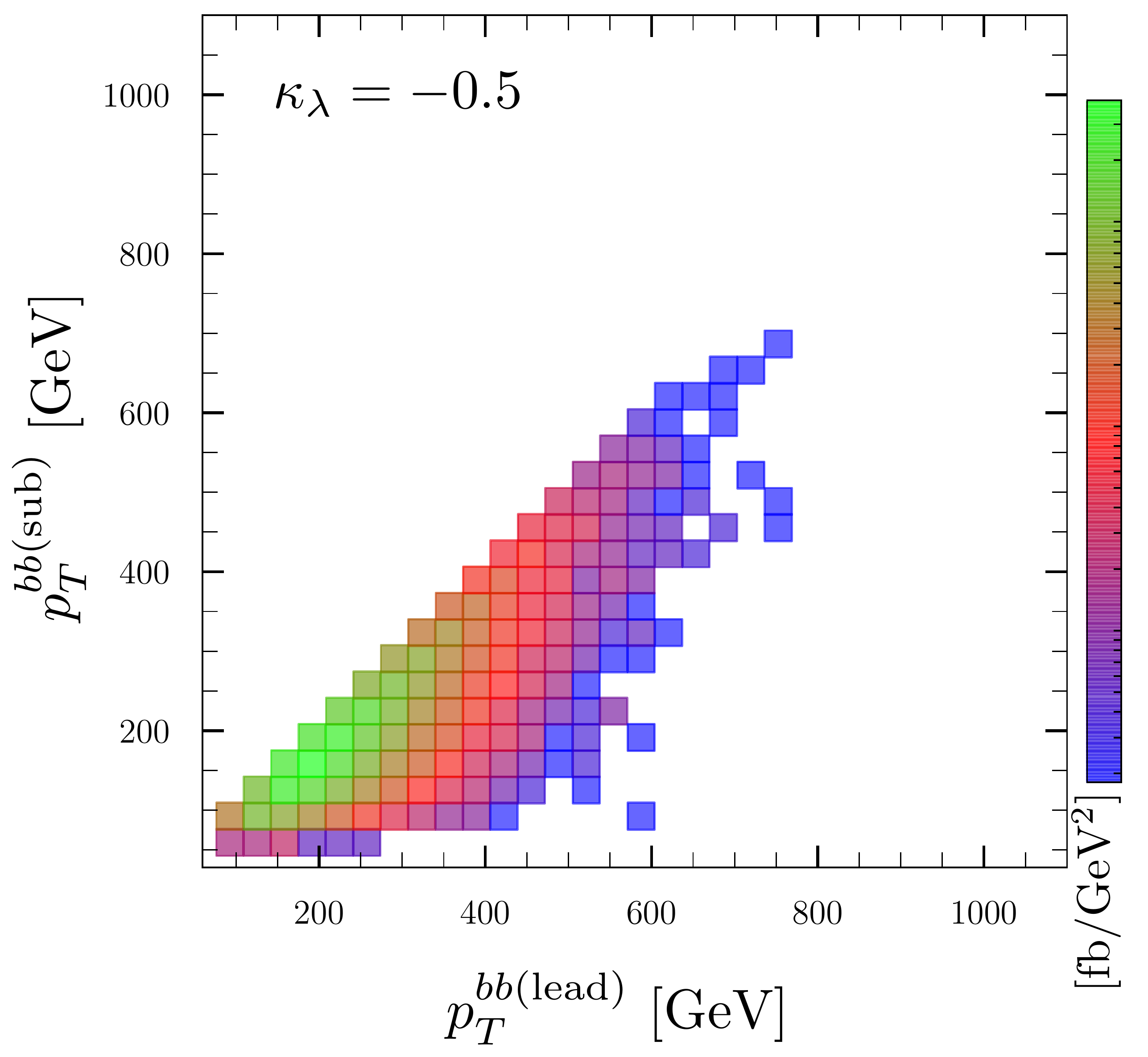}
    \includegraphics[height=180pt]{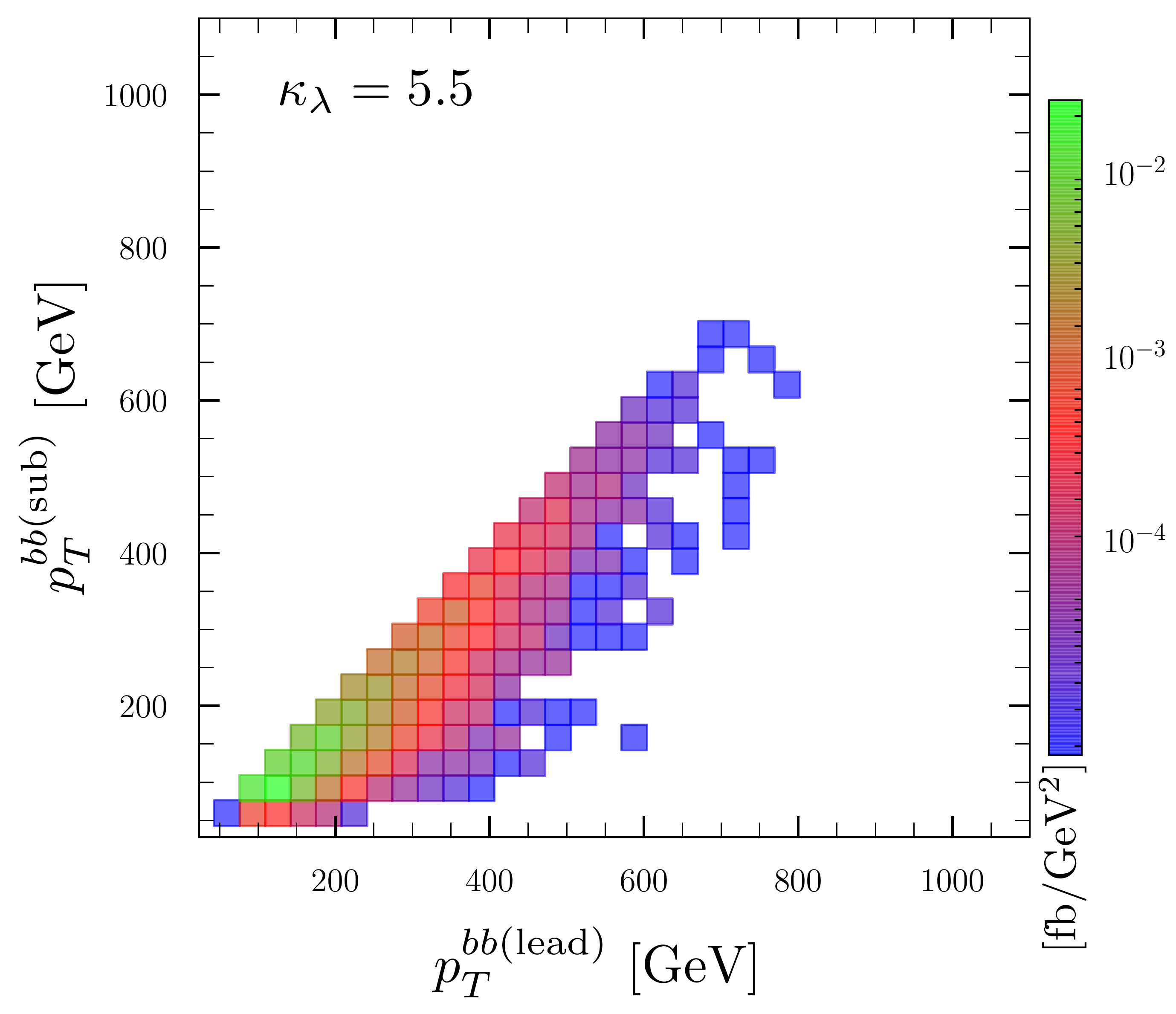}
    \caption{$d^2 \sigma/d p_T^{bb\rm (lead)} d p_T^{bb\rm (sub)} $
    in units of $\fb/{\rm GeV}^2$,
    where $p_T^{bb\rm (lead)}$ is the transverse momentum of the leading dijet
    and $p_T^{bb\rm (sub)}$ is that of the subleading dijet,
    in the SM (upper left), the VLQ-2HDM (upper right), 
    the SM with $\kp_\lm=-0.5$ (lower left) and
    $\kp_\lm=5.5$ (lower right). 
    }
    \label{fig:4b:ptpt}
\end{figure}

Motivated by the correlated bumps in the $p_T^{bb(\textrm{lead})}$ and $M_{4b}$ distributions, we study the double differential 
cross sections in some key variables.
In Fig.~\ref{fig:4b:ptpt}, 
we show one as a function of the transverse momentum 
of the leading dijet and the transverse momentum of the sub-leading dijet,
$d^2 \sigma/d p_T^{bb\rm (lead)} d p_T^{bb\rm (sub)}$,
in units of $\fb/\hbox{GeV}^2$.
We consider the SM (upper left), the VLQ-2HDM (upper right),
the SM with $\kp_\lm=-0.5$ (lower left) and
$\kp_\lm=5.5$ (lower right). 
The generic correlation 
of  $p_T^{bb\rm (lead)} \simeq p_T^{bb\rm (sub)}$,
originated from the back-to-back motion of two Higgs bosons,
is common for all four models.
The main difference is the {observable} kinematic \textit{area},
which is the largest for the VLQ-2HDM 
and the smallest for the case of $\kp_\lm=5.5$. 
In the region of
$p_T^{bb} > 300 \gev$,
only the VLQ-2HDM yields substantial number of events,
which is consistent with the parton-level result in Eq.~(\ref{eq:ratio:sigma}).
This unique feature is very useful for discriminating the VLQ-2HDM.

\begin{figure}
    \centering
    \includegraphics[height=180pt]{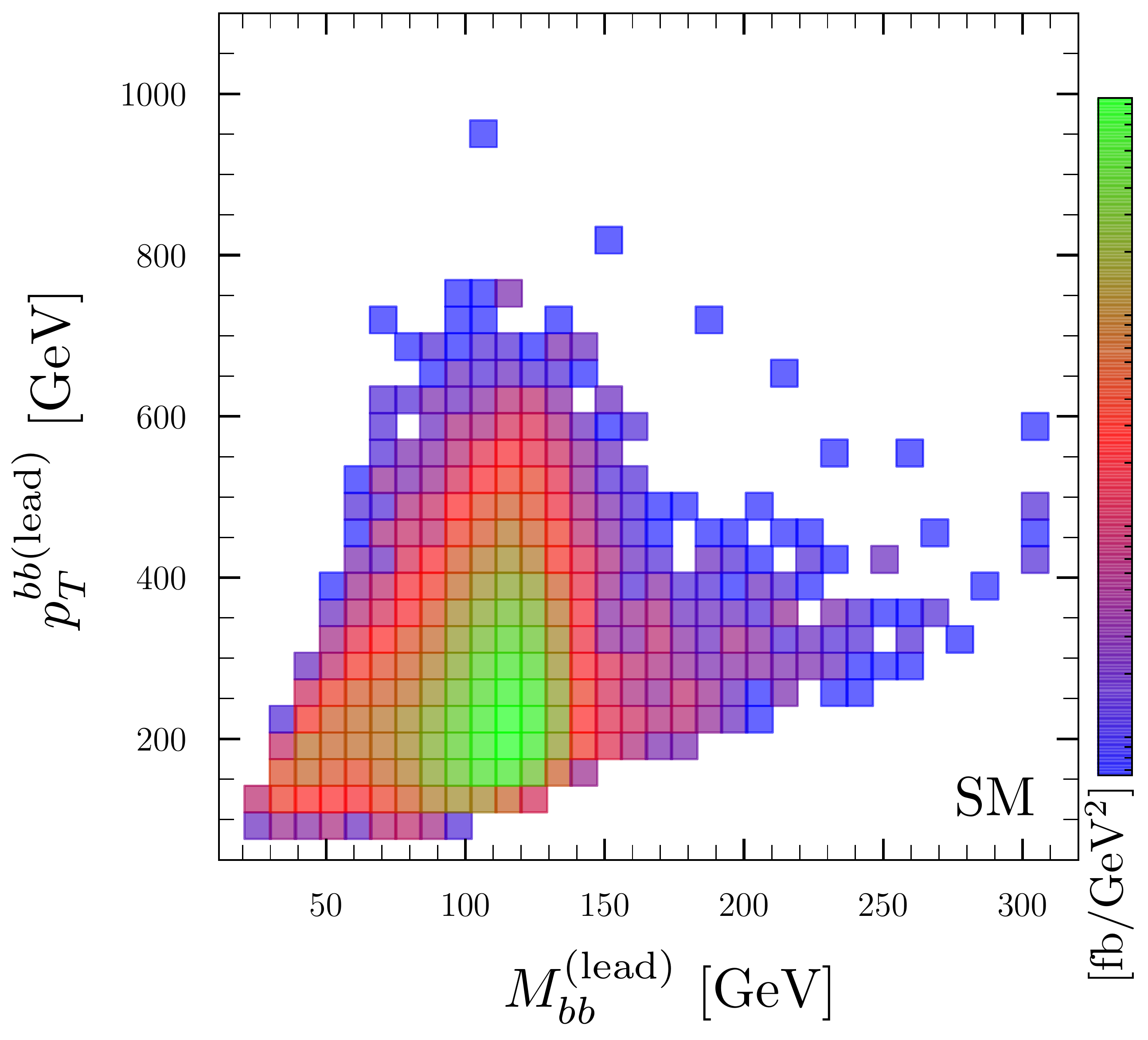}
    \includegraphics[height=180pt]{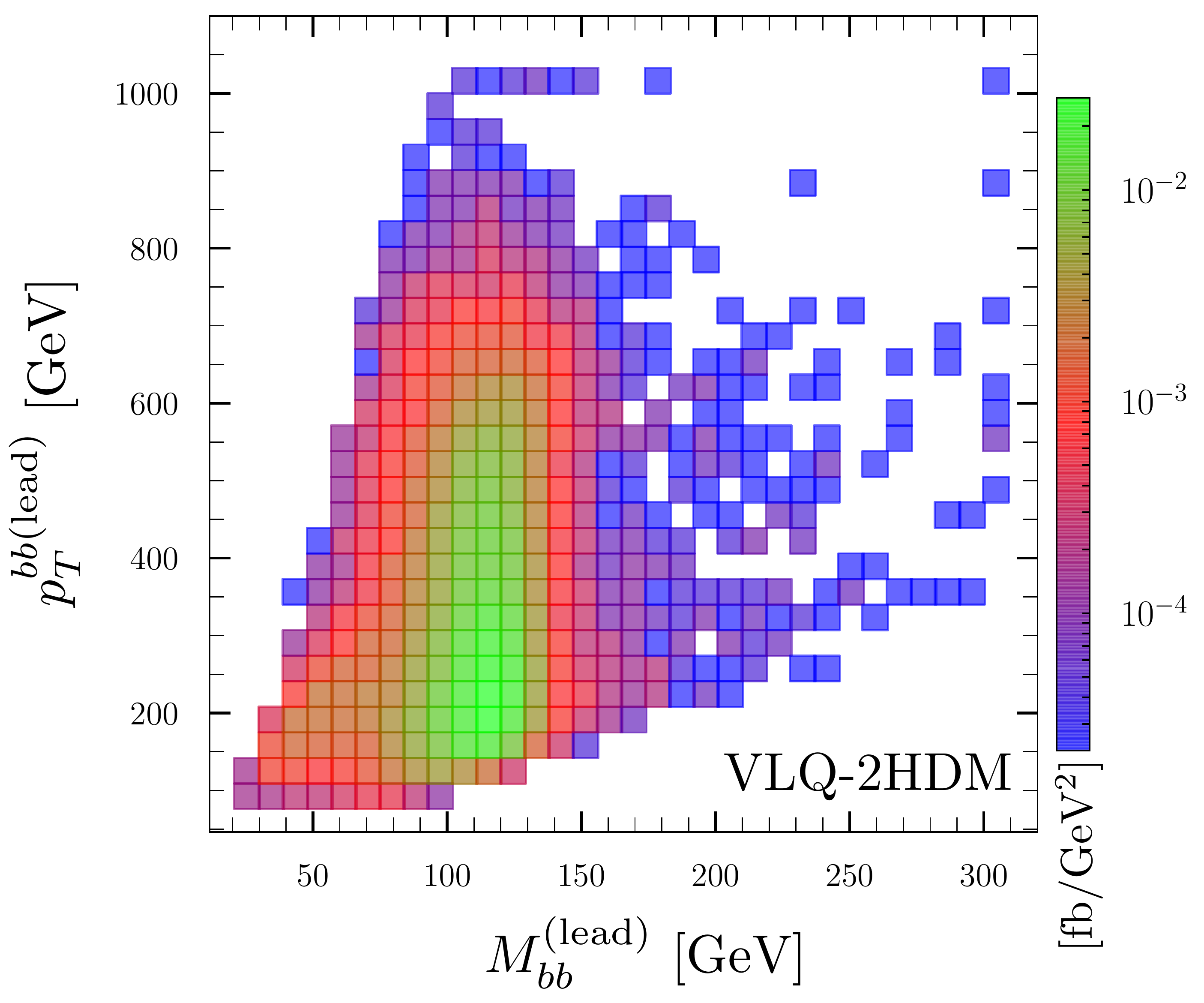}
    \\[12pt]
    \includegraphics[height=180pt]{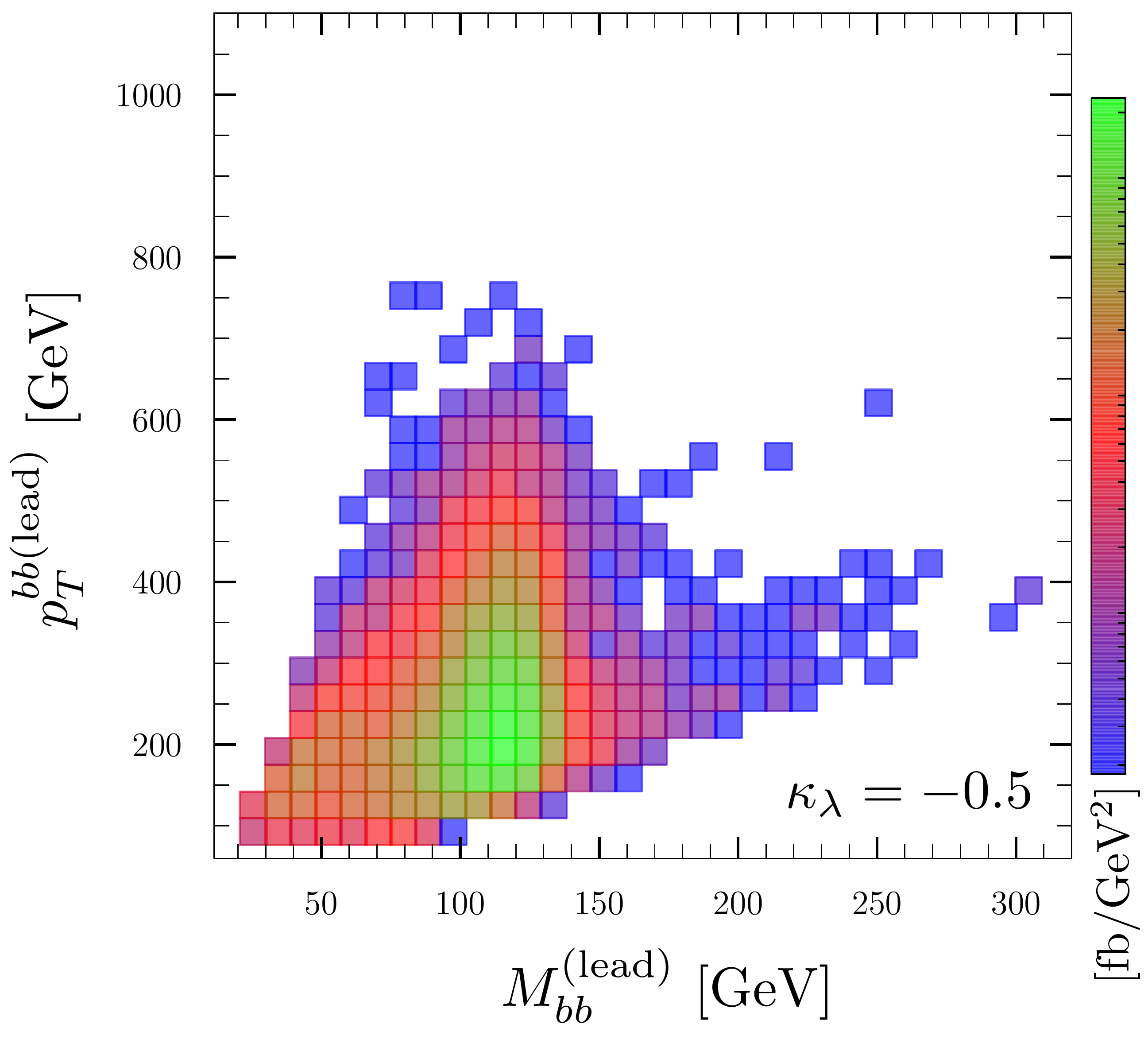}
    \includegraphics[height=180pt]{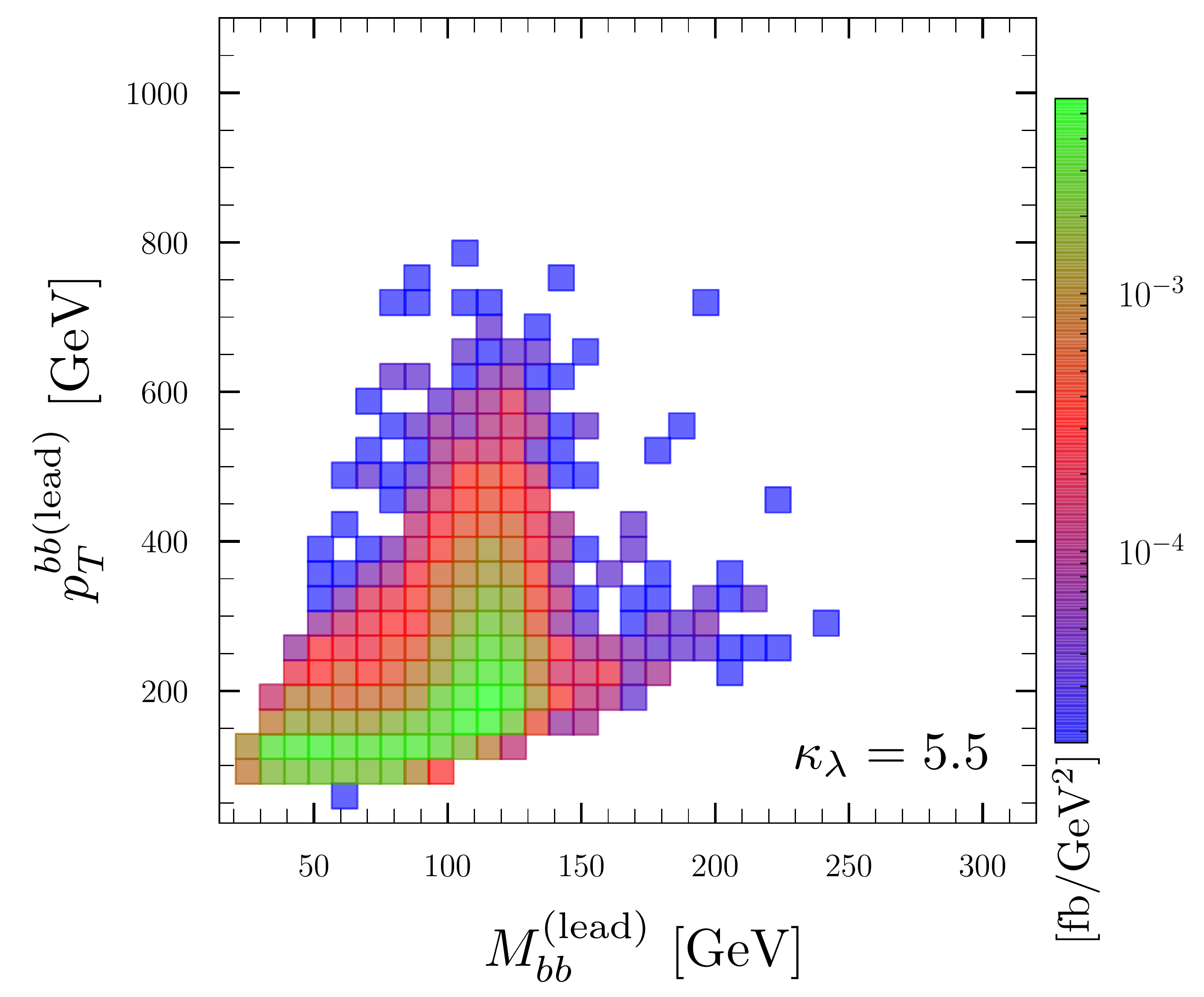}
    \caption{Same as Fig.~\ref{fig:4b:ptpt} but for 
    ${d^2 \sigma}/{d M_{bb}^{\rm (lead)} d p_T^{bb\rm (lead)} }$.}
    \label{fig:4b:mhpt}
\end{figure}

In Figure \ref{fig:4b:mhpt}, 
we display the double differential cross section in the invariant 
mass and transverse momentum of the leading dijet for the four models as shown in  Fig.~\ref{fig:4b:ptpt}.
The distributions
are well localized around the SM Higgs boson mass window ($M_{bb}^{\rm (lead)} \simeq m_h$)
in all of the models except for the case $\kp_\lm=5.5$
where a sizable number of events yield
$M_{bb}^{\rm (lead)}\lsim m_h$.
The LHC discovery prospect for $\kp_\lm=5.5$
is expected to be low, 
because the usual $m_h$ window cut removes 
a considerable part of the $\kp_\lm=5.5$ signal. 
Figure \ref{fig:4b:mhpt}
also shows that the correlation 
between $M_{bb}^{\rm (lead)} $ and $ p_T^{bb\rm (lead)}$
is very weak in all of the four models. 
Therefore, selecting events with high transverse momentum for the leading (or the subleading) dijet
does not alter the requirement on the Higgs boson mass windows.\footnote{Due to the different dynamics of the $4b$ from other decay modes of the di-Higgs process, 
the signal region requires $X_{hh} < 1.6$ where $
X_{hh} = \sqrt{\left(\frac{M_{bb}^{\rm (lead)} - 124\gev}{0.1 M_{bb}^{\rm (lead)}}\right)^2 + \left(\frac{M_{bb}^{\rm (sub)} - 115\gev}{0.1 M_{bb}^{\rm (sub)}}\right)^2}$.
Here a resolution of $10\%$ on the mass of the two dijets is assumed.} 
For the \textit{sub-leading} dijet, we find that the 
double differential cross section about its invariant mass and its transverse momentum 
shows a similar behavior as in Fig. \ref{fig:4b:mhpt}.

\begin{figure}
    \centering
    \includegraphics[height=180pt]{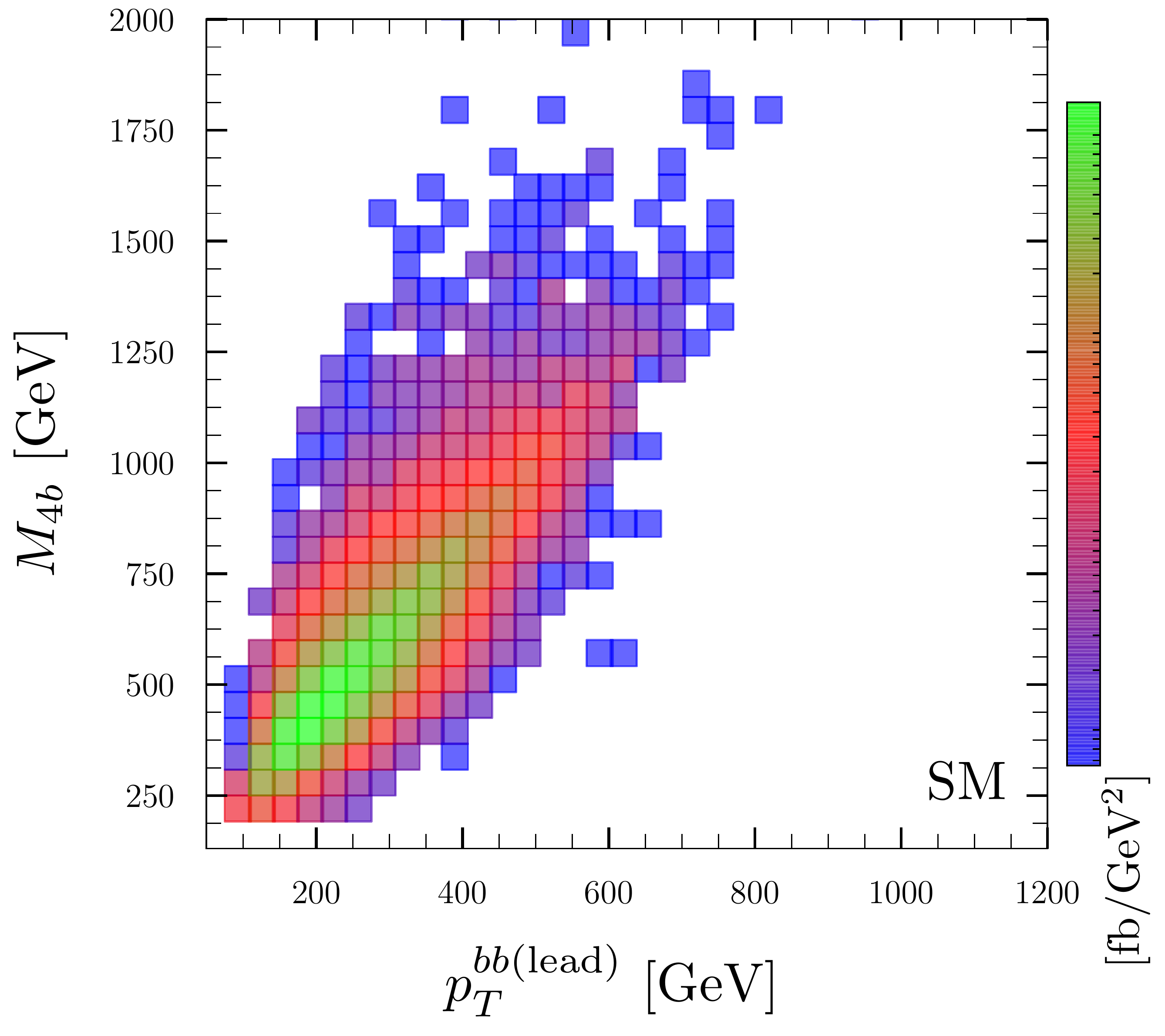}
    \includegraphics[height=180pt]{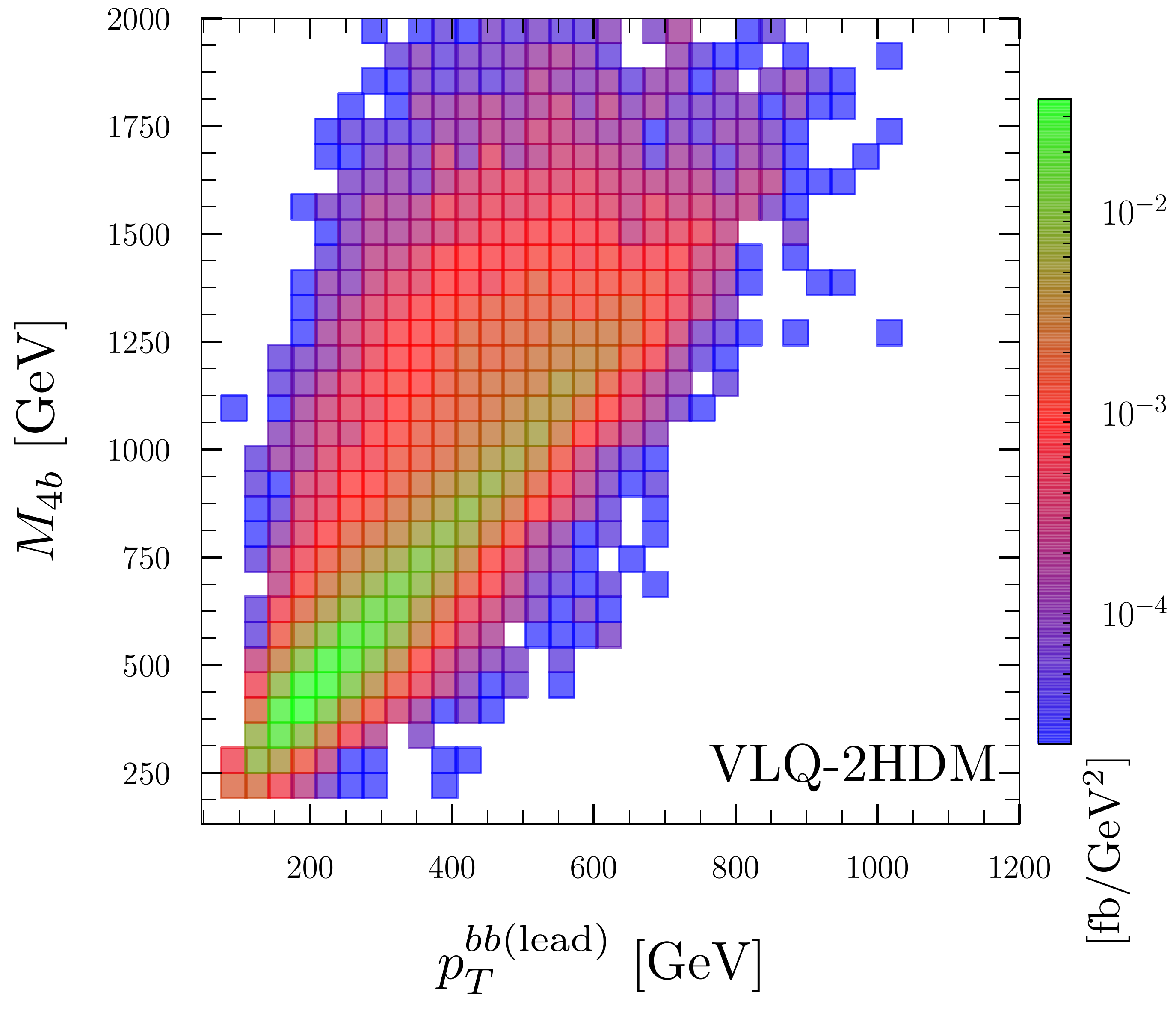}
    \\[12pt]
    \includegraphics[height=180pt]{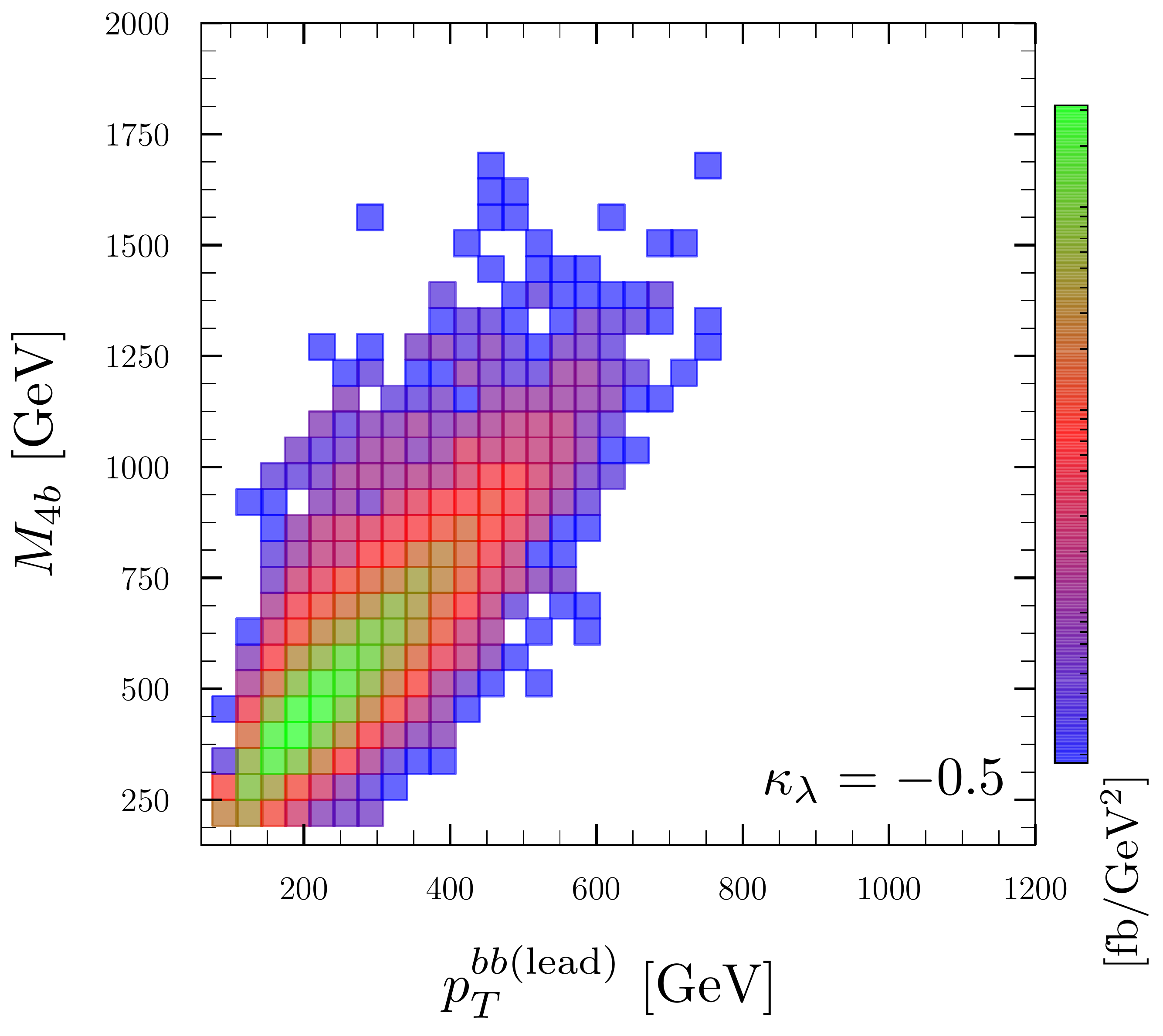}
    \includegraphics[height=180pt]{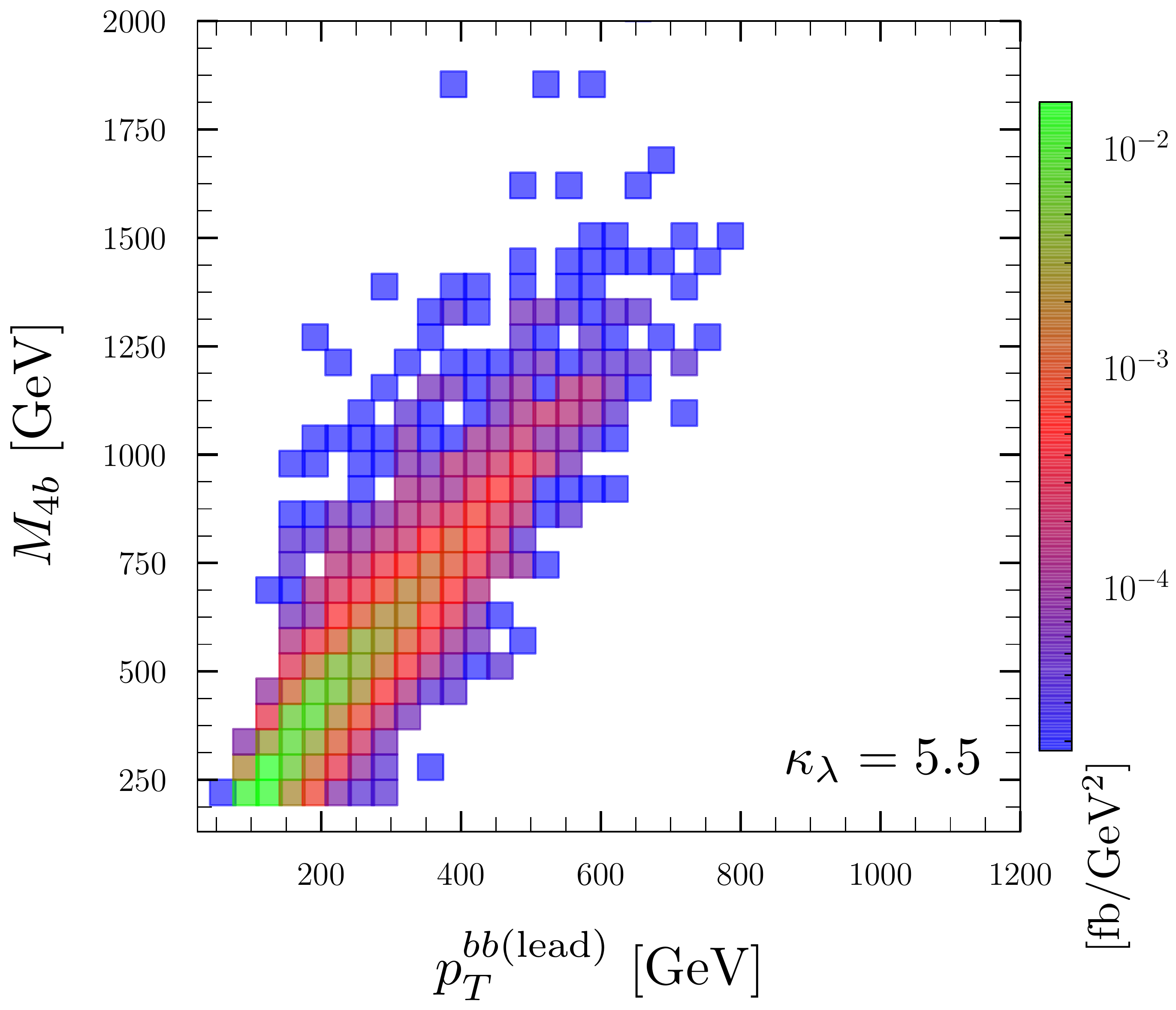}
    \caption{Same as Fig.~\ref{fig:4b:ptpt} but for 
    $d^2 \sigma/ d p_T^{bb\rm (lead)}d M_{4b} $.}
    \label{fig:4b:ptmhh}
\end{figure} 

Targeting two correlated bumps 
around $p_T^h \simeq M_1$ and $M_{hh}\simeq 2 M_1$ 
in the VLQ-2HDM,
we present the double differential cross section 
$d^2 \sigma/ d p_T^{bb\rm (lead)}d M_{4b}$
in Fig.~\ref{fig:4b:ptmhh}.
We observe a strong correlation 
along the line of $M_{4b} \simeq 2 p_T^{bb\rm (lead)}$
in all of the four models.
The unique feature of the VLQ-2HDM
is the extent of the \textit{observable} correlation
as well as its asymmetrical behavior toward the hard regions  in $M_{4b} $. 
High $M_{hh}$ cut along with high $p_T^h$ cut
will be one of the most sensitive probes for the VLQ-2HDM effects on the di-Higgs process.


\subsection{$\bb\rr$ final state}
For the analysis of the $\bb\rr$ final state of the di-Higgs process,
we follow the ATLAS reports~\cite{ATL-PHYS-PUB-2017-001,ATL-PHYS-PUB-2018-053}.
For the photon identification efficiency $\epsilon_\gamma$, 
we fit to the ATLAS simulation results and obtain 
the following dependence of $\epsilon_\gamma$
on the photon transverse momentum $p_T^\gm$:
\bea
\epsilon_\gamma =  0.888*\tanh\lf 0.01275 \frac{p_T^\gm}{{\rm GeV}}\ri\,.
\eea
The probabilities for a jet and an electron to fake a photon, the photon fake rates,   
are $P_{j\rightarrow \gamma}=5\times
10^{-4}$ and $P_{e\rightarrow \gamma}=2\%\,(5\%)$ in 
the barrel (endcap) region~\cite{ATL-PHYS-PUB-2017-001}.
For the $b$ tagging efficiency,  
we have fully adopted the dependence of $\epsilon_b$
on the transverse momentum and rapidity of the $b$-jet
in Fig.7(b) of Ref.~\cite{ATL-PHYS-PUB-2016-026}.
The misidentification probability of the charm quark jet as the $b$-jet, 
$P_{c \to b}$,
depends not only on the $b$-tagging efficiency 
but also
the transverse momentum and rapidity of the $c$-jet.
The $\epsilon_b$ dependence is incorporated by taking 
the 
multi-variate MV1 $b$-tagging algorithm
with
$P_{c \to b}\simeq 1/5$ for $\epsilon_b =0.7$ and 
$P_{c \to b}\simeq 1$ as $\epsilon_b\to 1$~\cite{ATL-PHYS-PUB-2015-022}.
The dependence of $P_{c \to b}$ on $p_T^c$ and $\eta^c$
is also included.
For the light-jet fake rate
as the $b$ jet, we take $P_{j \to b} = 1/1300$~\cite{ATL-PHYS-PUB-2017-001}.
The pile-up effects are not considered,
based on the reasonings in Ref.~\cite{Chang:2018uwu}.
The last consideration 
for a realistic analysis is the 
energy loss in the 
$b$ momentum reconstruction,
which is taken into account by the jet-energy scaling factor of
\bea
X_{E_b} = \sqrt{ \frac{(3.0 - 0.2|\eta_b|)^2 }{ p_{T}^b/{\rm GeV}} +1.27 }\,,
\eea
where the factor 1.27 is obtained by requiring a correct peak position at $M_{bb}=m_h$.
\begin{table}[t!]
\caption{\label{tab:selection} Sequence of the event preselection in $ hh \rightarrow b\overline{b}\gamma\gamma $ channel at the HL-LHC.}
\begin{center}
  {\renewcommand{\arraystretch}{1.3} 
   \begin{tabular}{|c|l| }
  \hline
    Sequence & Event Preselection at the HL-LHC\\
    \hline
    \hline
    1 & 
{Di-photon trigger condition:} \\ &
 $\geq $ 2 isolated photons with $p_T^\gm > 25$ GeV  and $|\eta_\gm| < 2.5$
\\
    \hline
    2 & $\geq $ 2 isolated photons with $p_T^\gm  > 30$ GeV,
$|\eta_\gm| < 1.37$ or $1.52 < |\eta_\gm| <2.37$, 
\\ & and $\Delta R_{j\gamma} > 0.4$ \\
    \hline
    3 & $\geq$ 2 $b$-jets with leading (sub-leading) $p_T^b > 40(30)$ GeV and $|\eta|<2.4$ \\
    \hline
    4 & $0.4 < \Delta R_{bb} < 2.0$ and $0.4 < \Delta R_{\gamma \gamma} < 2.0$ \\
    \hline
    \end{tabular}
    }
\end{center}
\end{table}

Referring to the ATLAS di-Higgs study in Ref.~\cite{ATL-PHYS-PUB-2017-001},
we take a sequence of the event preselection
for $gg\to hh \rightarrow b\overline{b}\gamma\gamma $ in Table \ref{tab:selection}.
We found that other preselections in Ref.~\cite{ATL-PHYS-PUB-2017-001}
are not useful for our signal.
In Table~\ref{tab:Eff},
we show the efficiencies of each sequence in four different models.
The efficiencies
are similar for the SM, the VLQ-2HDM,
and the SM with $ \kp_\lm=-0.5 $, 
being about $4$-$5\%$ at the final step.
In the case $ \kp_\lm=5.5 $, however, the efficiency dramatically  drops 
after the Selection-4, about a third of that in the other three models.
This is because the most events for the case $ \kp_\lm=5.5 $
are with $\Delta R_{\gamma\gamma} > 2.0 $ region like the main SM backgrounds~\cite{Chang:2018uwu}.
Even considering $\sg_\np/\sg_\sm\simeq 3$ for  $ \kp_\lm=5.5 $,
$\sg \times \br$ is about 80\% of the SM result after the Selection-4.
It is very challenging to probe at the HL-LHC. 

\begin{table}[h!]
\caption{
Cut flow efficiencies of four models for the di-Higgs 
process in the $\bb\rr$ final state
at the HL-LHC.
}
\label{tab:Eff}
\centering
\begin{tabular}{c||c|c|c|c}
\hline\hline
Sequence & ~~~SM~~~ & VLQ-2HDM & $ \kp_\lm=-0.5 $ & $ \kp_\lm=5.5 $ \\
\hline
1 & 27.60\% & 29.71\% & 25.19\% & 20.46\%\\ \hline
2 & 25.47\% & 25.88\% & 23.02\% & 18.12\% \\\hline
3 & 19.31\% & 18.35\% & 17.27\% & 12.86\% \\\hline
4 & \phantom{x}5.43\% & \phantom{x}4.78\% & \phantom{x}4.14\% & \phantom{x}1.51\% \\
\hline\hline
\end{tabular}
\end{table}

\begin{figure}
    \centering
    \includegraphics[height=160pt]{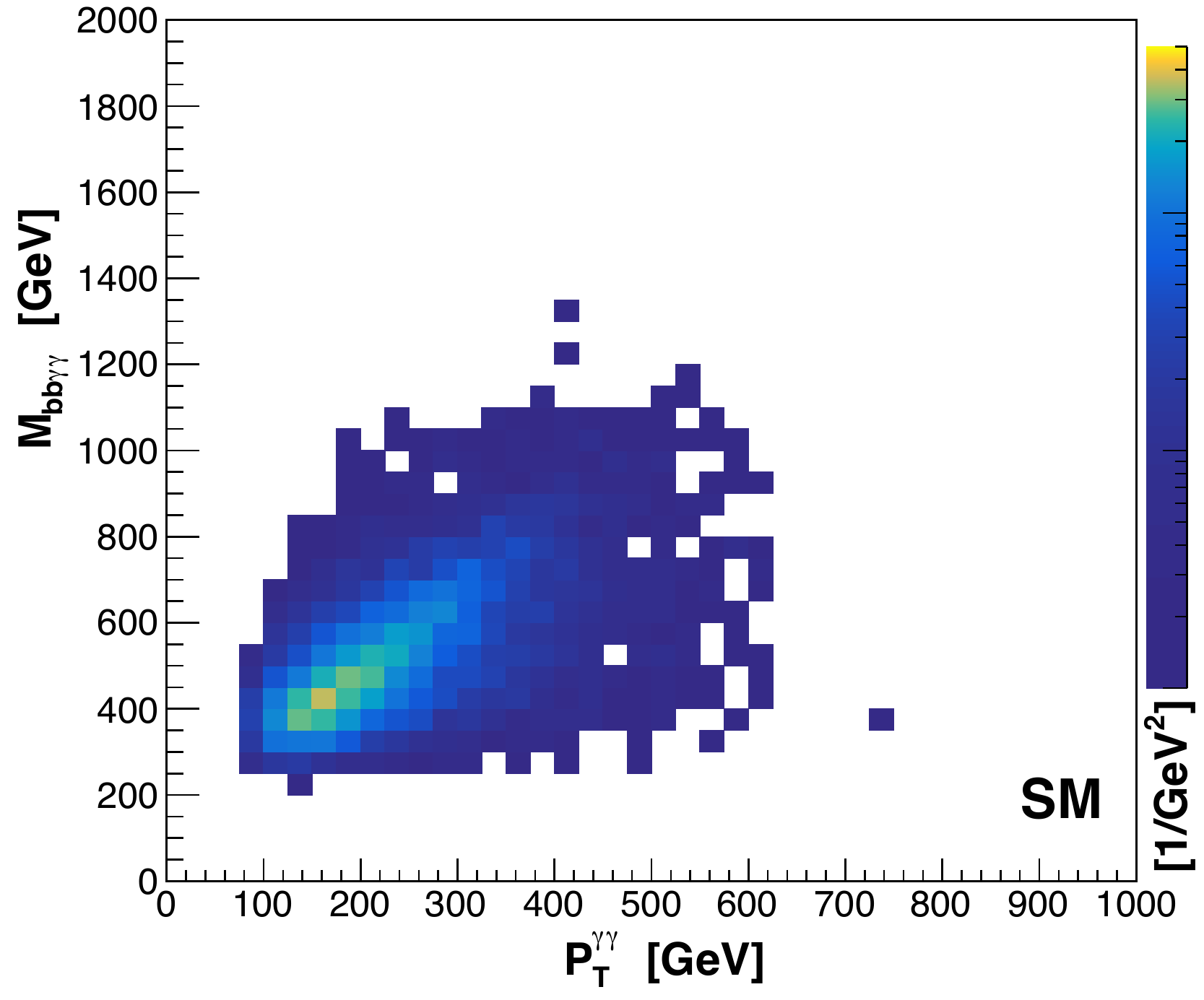}
    \includegraphics[height=160pt]{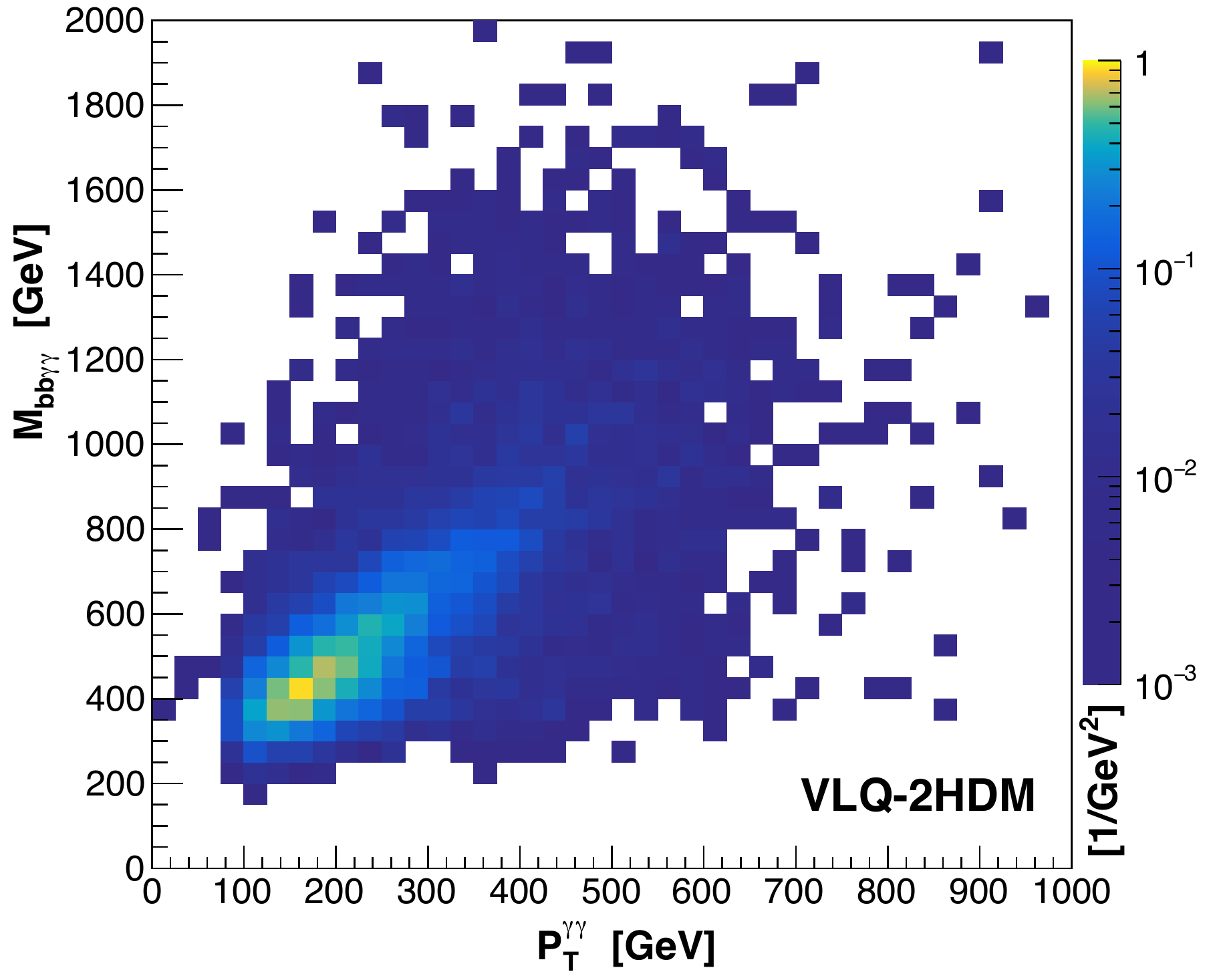}
    \\[12pt]
    \includegraphics[height=160pt]{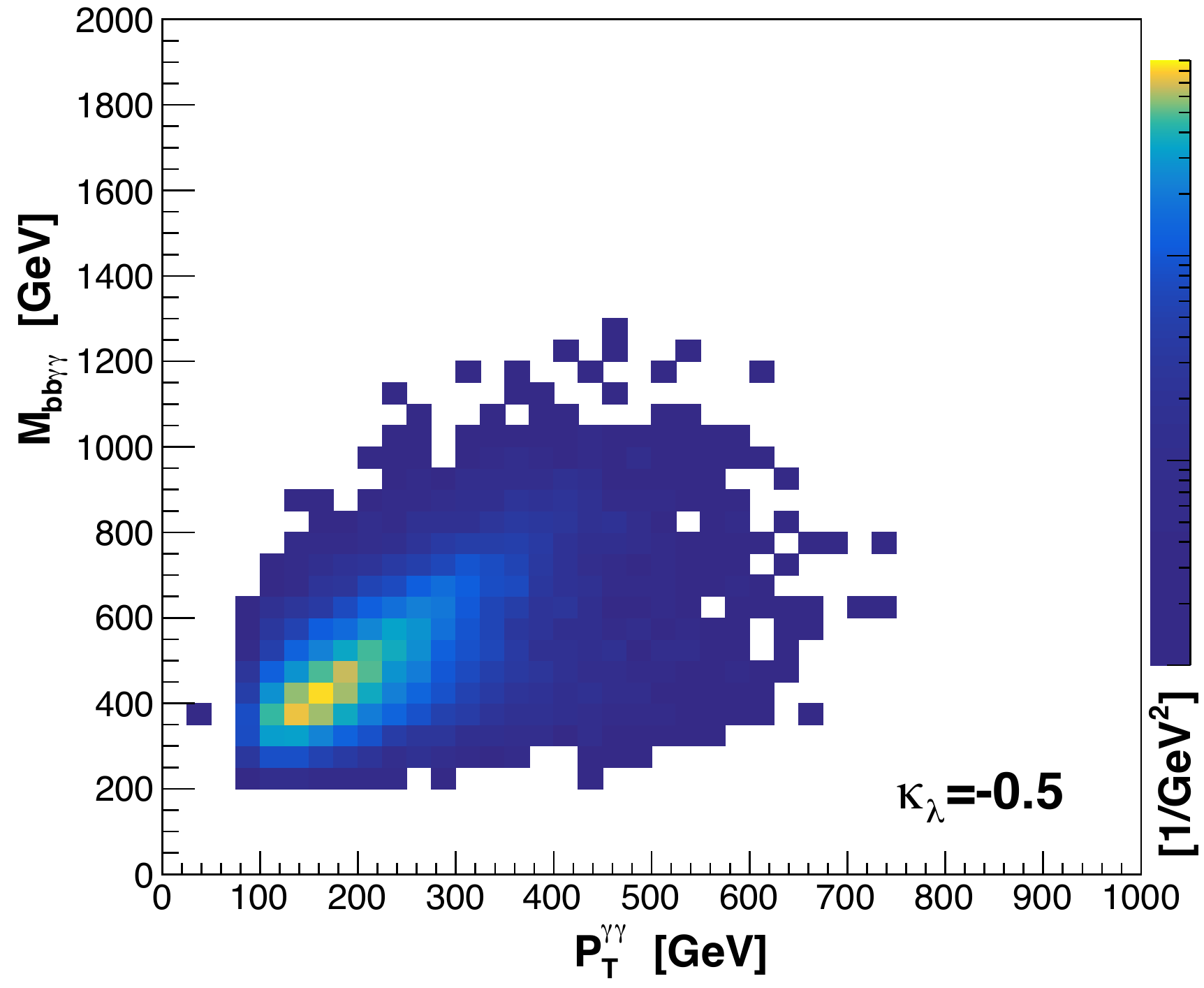}
    \includegraphics[height=160pt]{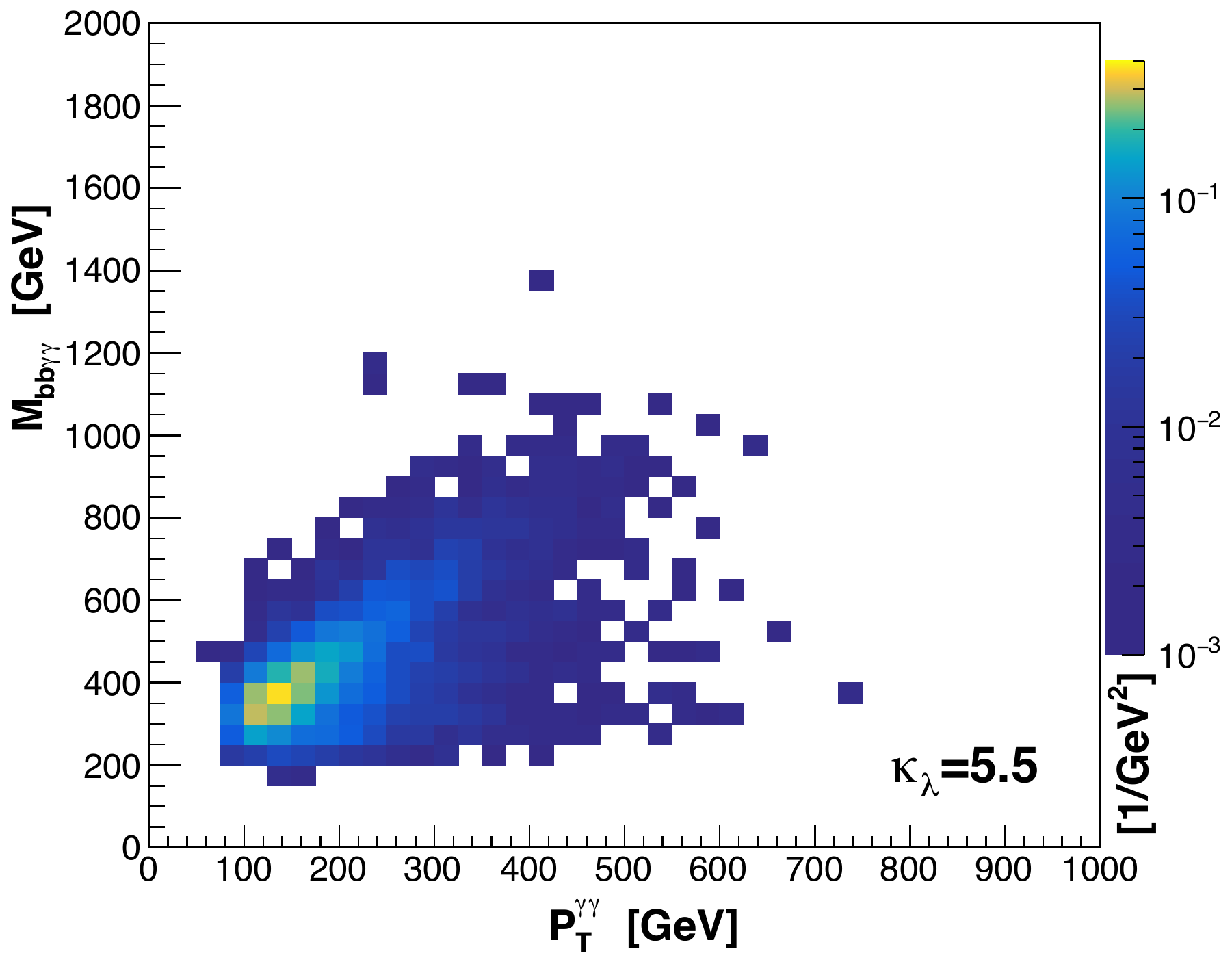}
    \caption{The distribution of the number of events 
    versus the transverse momentum of the {di-photon} and the invariant mass of $\bb\rr$ 
    for the $\bb\rr$ final state of the di-Higgs process
    at the HL-LHC.
    We consider the SM (upper left), the VLQ-2HDM (upper right),
    the SM with $\kp_\lm=-0.5$ (lower left), and $\kp_\lm=5.5$ (lower right).
    }
    \label{fig:bbaa:ptmhh}
\end{figure} 

In Fig.~\ref{fig:bbaa:ptmhh},
we present $d^2 N/d p_T^\rr d M_{\bb\rr}$,
the distribution of the number of events 
versus the transverse momentum of the di-photon and the invariant mass of $\bb\rr$,
in the SM (upper left), the VLQ-2HDM (upper right),
    the SM with $\kp_\lm=-0.5$ (lower left), and $\kp_\lm=5.5$ (lower right).
Since the $\bb\rr$ final state has an extremely small cross section,
we show the distribution of the number of events 
corresponding to the total integrated luminosity of $3000\ifb$.
The overall characteristics are very similar to those in the $\bb\bb$ final state:
there is a strong correlation along the line $ M_{\bb\rr} \simeq 2 p_T^\rr$
in all of the four models;
the VLQ-2HDM yields the widest spread up to high $p_T^\rr$ and $ M_{\bb\rr}$;
the $\kp_\lm=5.5$ case prefers small $p_T^\rr$ and  $ M_{\bb\rr}$, 
compared with the other models.
If we count the bins with $d^2 N/d p_T^\rr d M_{\bb\rr} > 1/{\rm GeV}^2$,
however,
it is very difficult to see the difference among different NP models. 
Moreoever, the isolation condition, $ \Delta R_{\gamma \gamma}, \Delta R_{bb} > 0.4 $, also restricts the power to detect high $ p_T^\rr, p_T^\bb $ regions in the $\bb\rr$ final state.
In summary, the $\bb\rr$ final state plays a
complementary role in observing the di-Higgs process,
but not appropriate for the deeper study of NP.

\section{Conclusions}
\label{sec:conclusion}

With the aim of disentangling different NP contributions
to the di-Higgs process from gluon fusion,
we have studied the phenomenological characteristics of the
kinematical distributions, 
focusing on the double differential cross sections.
For illustration purpose, we assume that the NP effects would
 first appear 
 in the total
  production cross section, being three times as large as the SM expectation.
Since we can easily identify resonant di-Higgs production,
we concentrated on the non-resonant NP effects from
non-SM Higgs trilinear couplings ($\kp_\lm=-0.5$ or $\kp_\lm=5.5$)
and the new colored fermions running in the loop.
For the latter,
we need a concrete NP model for a comprehensive 
study since
new quarks,
which enhance the di-Higgs production rate,
should similarly act in the single-Higgs production.

In this work, we have studied a type-II 2HDM with vectorlike quarks,
called the VLQ-2HDM.
The electroweak oblique parameters remain almost the same as in the SM
by adopting an ansatz that guarantees a vanishing $\hat{T}$,
called the zero-$\hat{T}$ ansatz: see Eq.~(\ref{eq:ansatz}).
We analytically calculated the new form factors from the VLQs. 
In order to show the role of the Higgs-fermion-fermion couplings 
in breaking the correlation between the di-Higgs 
and single-Higgs processes,
we considered the alignment limit and the exact wrong-sign (EWS) limit.
In the alignment limit,
both up-type and down-type VLQs 
have the same-sign couplings to the Higgs boson,
so that their contributions to the triangle diagrams of the di-Higgs process
are constructive to each other.
Moreover they are strongly correlated with the VLQ contributions
to the box diagrams.
As the single-Higgs process has the same triangle diagrams,
we cannot accommodate $\sg_\np/\sg_\sm (gg\to hh) \simeq 3$
and $\sg_\np/\sg_\sm (gg\to h) \simeq 1$ simultaneously:
the maximum increase of the di-Higgs production rate allowed by
the observed $\kp_g$ is only $20\%$.
In the EWS limit, however, the
down-type and up-type VLQs have opposite sign Yukawa couplings, thus
yielding a considerable cancellation between their contributions
to single-Higgs production.
The box diagrams do not have this kind of cancellation 
because their amplitudes are proportional to the square of the Higgs-fermion-fermion coupling.
Significant enhancement of the total production cross section of the di-Higgs process
is feasible 
in the EWS limit, where we took a benchmark point.

First at parton level, 
we calculated the kinematic distributions for the $gg \to hh$ process
in three NP models, $\kp_\lm=-0.5$, $\kp_\lm=5.5$, and the VLQ-2HDM.
Although they have almost the same total production cross section
of $\sg_\np/\sg_\sm (gg \to hh)\simeq 3$,
the $M_{hh}$ and $p_T^h$ distributions show quite significant differences.
The $\kp_\lm=-0.5$ model yields similar distribution shapes to the SM results.
In the $\kp_\lm=5.5$ model, 
both distributions apparently shift toward
low $M_{hh}$ or $p^h_T$ region such that 
  the peak position moves about $100\gev$.
This feature
makes the $\kp_\lm=5.5$ model very challenging to probe at the LHC,
because the SM backgrounds to the di-Higgs process such as $4b$, $\bb c\bar{c}$,
and $\ttop$ are populated in the low $p_T^h$ region.
The VLQ-2HDM showed its unique and distinctive features
in the $M_{hh}$ and $p_T^h$ distributions,
benefiting from the VLQ threshold effects.
At parton level,
we could clearly see the bumps around $M_{hh}\simeq 2 M_1$ and $p_T^h\simeq M_1$,
where $M_1$ is the lightest VLQ mass.
Moreover, the bumps of the threshold origin from heavy VLQs
naturally lift up the kinematic distributions of
$M_{hh}$ and $p_T^h$ into high regions.
The doubly high region, with high $M_{hh}$ and high $p_T^h$,
can be the exclusive territory of the VLQ-2HDM for the di-Higgs process.

We also have completed the analysis with full collider simulations
for the di-Higgs signals
in the VLQ-2HDM, the SM, the SM with $\kp_\lm=-0.5$, and with $\kp_\lm=5.5$.
Two final states, $\bb\bb$ and $\bb\rr$
of the decays of the Higgs-boson pair, were studied.
Fortunately,
many characteristic features at the parton-level calculation
survived even after parton showering, hadronization, and detector simulations.
The bump structures in the distributions of $M_{hh}$ and $p_T^h$,
though being smeared a little bit,
are maintained, 
  and the positions of the peaks roughly stay at the same place.
Motivated by the 
correlation of the bumps in $M_{hh}$ and $p_T^h$ distributions,
we studied various double differential cross sections.
In the $\bb\bb$ final state, 
we first found that any selection on the transverse momentum of the leading 
(or the sub-leading) dijets since a Higgs boson candidate barely alters its invariant mass.
The smoking-gun signature appears in 
$d^2 \sg/d M_{hh}\, dp_T^h$.
All four models
showed a strong correlation along the line of $M_{hh} \simeq 2 p_T^h$,
which is also useful to search for the SM di-Higgs process itself.
Distinguishing the VLQ-2HDM from other NP models
is possible in the $\bb\bb$ final state
as the observable correlation line of $M_{hh} \simeq 2 p_T^h$
is the longest, 
extending far toward high $p_T^h$ region:
the case $\kp_\lm=5.5$ has the shortest.
However, the $\bb\rr$ final state has too small signal rate,
not appropriate to
see the difference among the NP models.
In summary,
we expect that our observation of the correlation between $M_{hh}$ and $ p_T^h$
distributions
for disentangling the NP effects on the di-Higgs process can help the NP search at the HL-LHC.

\acknowledgments
K.C. was supported by the MoST of Taiwan 
under Grant No. MOST-107-2112-M-007-029-MY3.
The work of AJ and JS is supported 
by the National Research Foundation of Korea, Grant No. NRF-2019R1A2C1009419.
The work of YWY was supported by the National Research Foundation of Korea(NRF) grant
funded by the Korea government(MSIT) (No. 2019R1I1A1A01064113)


\end{document}